\documentclass[12pt]{article}
\usepackage{amsthm,amsfonts,amsmath,amssymb,mathrsfs,url}
\usepackage[usenames,dvipsnames]{color}

\title{Multi-Time Wave Functions for\\ Quantum Field Theory}

\author{
S\"oren Petrat\footnote{Mathematisches Institut, 
	Ludwig-Maximilians-Universit\"at, Theresienstr. 39, 80333 M\"unchen, Germany.
	E-mail: petrat@math.lmu.de}\ \ and
Roderich Tumulka\footnote{Department of Mathematics,
     Rutgers University,
     110 Frelinghuysen Road, Piscataway, NJ 08854-8019, USA.
     E-mail: tumulka@math.rutgers.edu}
}
\date{January 24, 2014}

\theoremstyle{plain}
\theoremstyle{plain}\newtheorem{ass}{Assertion}
\theoremstyle{plain}
\theoremstyle{definition}

\newcommand{\be}{\begin{equation}}
\newcommand{\ee}{\end{equation}}
\newcommand{\Hilbert}{\mathscr{H}}
\newcommand{\conf}{\mathcal{Q}}

\newcommand{\RRR}{\mathbb{R}}

\newcommand{\CCC}{\mathbb{C}}

\newcommand{\scp}[2]{\langle #1|#2 \rangle}
\newcommand{\Bscp}[2]{\Bigl\langle #1\Big|#2 \Bigr\rangle}

\newcommand{\vy}{\boldsymbol{y}}
\newcommand{\valpha}{\boldsymbol{\alpha}}
\newcommand{\vx}{\boldsymbol{x}}

\newcommand{\sB}{\mathscr{B}}

\newcommand{\sM}{\mathscr{M}}

\newcommand{\sS}{\mathscr{S}}

\newcommand{\free}{\mathrm{free}}
\newcommand{\ip}{\mathrm{ip}} 

\newcommand{\Fock}{\mathscr{F}}
\newcommand{\Hd}{\mathcal{H}}
\newcommand{\inter}{\mathrm{int}}

\newcommand{\sA}{\mathscr{A}}

\DeclareMathOperator{\Anti}{Anti}
\DeclareMathOperator{\Sym}{Sym}
\DeclareMathOperator{\Ann}{Ann}
\DeclareMathOperator{\Cr}{Cr}

\newcounter{remarks}
\begin{document}
\maketitle

\begin{abstract}
Multi-time wave functions such as $\phi(t_1,\vx_1,\ldots,t_N,\vx_N)$ have one time variable $t_j$ for each particle. This type of wave function arises as a relativistic generalization of the wave function $\psi(t,\vx_1,\ldots,\vx_N)$ of non-relativistic quantum mechanics. We show here how a quantum field theory can be formulated in terms of multi-time wave functions. We mainly consider a particular quantum field theory that features particle creation and annihilation. Starting from the particle--position representation of state vectors in Fock space, we introduce multi-time wave functions with a variable number of time variables, set up multi-time evolution equations, and show that they are consistent. Moreover, we discuss the relation of the multi-time wave function to two other representations, the Tomonaga--Schwinger representation and the Heisenberg picture in terms of operator-valued fields on space-time. In a certain sense and under natural assumptions, we find that all three representations are equivalent; yet, we point out that the multi-time formulation has several technical and conceptual advantages.

\medskip

\end{abstract}

\newpage
\tableofcontents

\begin{quotation}
{\it ``I, therefore, think that a good theoretical physicist today might find 
it useful to have a wide range of physical viewpoints and mathematical 
expressions of the same theory (for example, of quantum electrodynamics) 
available to him.''} \hfill Richard Feynman, Nobel lecture 1965 \cite{Fey65}
\end{quotation}

\section{Introduction}

The most naively obvious way of turning a wave function 
\be
\psi(t,\vx_1,\ldots,\vx_N)
\ee
of non-relativistic quantum mechanics ($\vx_j\in\RRR^3$) into a relativistic object is to replace it by a \emph{multi-time wave function}
\be\label{phi}
\phi(t_1,\vx_1,\ldots,t_N,\vx_N)\,,
\ee
that is, by a function of $N$ space-time points $x_k=(x_k^0,x_k^1,x_k^2,x_k^3)=(x^0_k,\vx_k)=(t_k,\vx_k)$. (We set $c=1=\hbar$ throughout.) We explore in this paper how this  approach can be applied in quantum field theory (QFT), where the number $N$ of particles is not fixed. We take for granted that the Hilbert space is a tensor product of Fock spaces, and that state vectors possess a particle--position representation, which is a function on a configuration space of a variable number of particles (a ``Fock function''), such as
\be\label{confFock}
\conf = \bigcup_{N=0}^\infty (\RRR^3)^{N} = \Gamma(\RRR^3)
\ee
with the notation
\be\label{Gammadef}
\Gamma(S):=\bigcup_{N=0}^\infty S^{N}\,.
\ee

For electrons and positrons, we know that such a representation exists. There are unsolved problems concerning the position representation of photons (see, e.g., \cite{LP:1930,BB,SZ}), and we leave aside the problem whether a particle--position representation exists for them. However, it is widely agreed that a 1-photon quantum state can be mathematically described by a (complexified) Maxwell field, so it does possess some kind of position representation, which is enough to enable the application of the kind of multi-time wave functions considered in this paper.

A multi-time version of a wave function on the configuration space \eqref{confFock} would be a function $\phi$ on either $\Gamma(\RRR^4)=\cup_{N=0}^\infty (\RRR^4)^N$ or the set of all \emph{spacelike configurations},\footnote{We use the symbol $i$ in two different meanings, for the unit imaginary number and as a particle label $i\in\{1,\ldots,M\}$. It should always be clear which one is meant.}
\be\label{Sdef}
\sS = \bigcup_{N=0}^\infty \Bigl\{ (x_1,\ldots,x_N)\in(\RRR^4)^N:  
\:\: \forall i\neq j\in\{1\ldots N\}: x_i \sim x_j \text{ or } x_i=x_j \Bigr\}\,,
\ee 
where $x\sim y$ means that $x$ is spacelike separated from $y$, i.e., $(x^0-y^0)^2-\|\vx-\vy\|^2<0$. Of such a function $\phi$ (a ``multi-time Fock function'') we say that it has a variable number of time variables, just like a Fock function on $\Gamma(\RRR^3)$ has a variable number of space variables. Configurations with two or more particles at the same space-time points will be called \emph{collision configurations}; note that we do not exclude them from $\sS$.
Bloch \cite{bloch:1934} has first argued that multi-time wave functions should be defined on \emph{spacelike} configurations only. We share this view on the grounds that, for theories involving particle creation and annihilation, the multi-time equations are inconsistent on $\Gamma(\RRR^4)$ but consistent on $\sS$; this conclusion will be supported by results of this paper (see Assertion~\ref{ass:consistency} in Section~\ref{sec:MT-em-ab}) and of \cite{pt:2013b}.

In \cite{pt:2013a} we discuss multi-time wave functions for a fixed number $N$ of particles and provide proofs of the need for consistency conditions. (These are conditions that a system of multi-time evolution equations must satisfy to possess solutions for all initial conditions; see also Remark~\ref{rem:consistency} in Section~\ref{sec:MT-em-ab} below. While multi-time equations for non-interacting particles are always consistent, it is challenging to set up consistent multi-time equations with interaction.) The main result of \cite{pt:2013a} is that interaction potentials (given by multiplication operators) always lead to inconsistency. We conclude that interaction has to be implemented by other means, namely by creation and annihilation of particles, as is done in QFT. That was the main motivation for this paper. Our key result (Assertion~\ref{ass:consistency}) is that, indeed, multi-time equations with interaction by creation and annihilation of particles are consistent. As far as we know, this work provides the first consistent multi-time model with interaction that can reproduce the well-known predictions of quantum (field) theory.

\subsection{Overview}

In this work, we study the use of multi-time wave functions for a model QFT in which there are two particle species, say $x$-particles and $y$-particles, and the $x$-particles can emit and absorb $y$-particles. (Henceforth, this model QFT is called \emph{the emission--absorption model}.) Starting from a standard single-time formulation of the model in the particle--position representation, we set up suitable multi-time evolution equations, \eqref{multi12}, and derive (by formal calculation without mathematical rigor) that they are consistent. A rigorous consistency proof of a version of the equations with an ultraviolet (UV) cut-off will be provided in \cite{pt:2013b}. These multi-time equations, \eqref{multi12}, are the key equations of this paper. They are remarkably simple and elegant; formulated in terms of the manifestly covariant object $\phi$, they would be manifestly covariant equations, see \eqref{multi34}, if the specific particle creation and annihilation terms that we use did not prefer one Lorentz frame. The equations \eqref{multi12} are a system of coupled partial differential equations of first order, and are essentially of the form 
\be\label{phiHj}
i\frac{\partial}{\partial t_j}\phi(t_1,\vx_1,\ldots,t_N,\vx_N) = H_j \phi(t_1,\vx_1,\ldots,t_N,\vx_N)
\ee
at every configuration $(t_1,\vx_1,\ldots,t_N,\vx_N)\in\sS$, where $H_j\phi$ involves also the $N+1$- and the $N-1$-particle sector of the multi-time Fock function $\phi$. Note that the number of equations varies from sector to sector of configuration space-time $\sS$, unlike in most previous works on multi-time equations. In contrast to Dirac \cite{dirac:1932} and Dirac, Fock, and Podolsky \cite{dfp:1932}, who suggested to take the wave function in a particle representation for fermions and in a field representation for bosons, we will use a particle representation for both fermions and bosons. We obtain that the multi-time wave function shares the permutation symmetry of the wave function of quantum mechanics (i.e., it is symmetric against permutation of identical bosons and anti-symmetric against permutation of identical fermions), except that, while one interchanges \emph{space points} in ordinary quantum mechanics, one interchanges \emph{space-time points} in the multi-time formulation (as already suggested by Marx \cite{Marx:1972}). 

We describe an analogous multi-time formulation of another model QFT, a simple model of electron--positron pair creation and annihilation, in \cite{pt:2013d}.

Furthermore, we examine in this paper the translation between multi-time wave functions and two better-known covariant formulations of QFT: the Tomonaga--Schwinger equation and the approach of operator-valued fields. 

In the Tomonaga--Schwinger approach \cite{tomonaga:1946,KTT:1947a,KTT:1947b,schwinger:1948}, one attributes a vector $\tilde\psi_\Sigma$ in a fixed Hilbert space $\tilde\Hilbert$ to every spacelike hypersurface $\Sigma\subset\RRR^4$. (Throughout this paper, we simply say ``spacelike hypersurface'' for ``spacelike Cauchy hypersurface.'') The Tomonaga--Schwinger approach employs the interaction picture (as distinct from the Schr\"odinger picture and the Heisenberg picture) and provides an equation for how $\tilde\psi_{\Sigma}$ changes as we continuously change $\Sigma$, the \emph{Tomonaga--Schwinger equation}. It reads
\be\label{TS}
i\bigl(\tilde\psi_{\Sigma'}-\tilde\psi_\Sigma\bigr) = \biggl( \int_{\Sigma}^{\Sigma'} \!\!\!\! d^4 x \, \Hd_I(x)\biggr)\, \tilde\psi_\Sigma
\ee
for infinitesimally neighboring spacelike hypersurfaces $\Sigma,\Sigma'$. Here, $\int_{\Sigma}^{\Sigma'} d^4x$ means the integral over the oriented (infinitesimal) 4-dimensional volume enclosed between $\Sigma$ and $\Sigma'$, and $\Hd_{I}(x)$ is the interaction Hamiltonian density on $\tilde\Hilbert$ in the interaction picture. The Tomonaga--Schwinger equation \eqref{TS} and the multi-time equation \eqref{phiHj} have in common that certain commutator conditions on the Hamiltonians $\Hd_I(x)$ (respectively, on the $H_j$) are needed to ensure consistency (or integrability) of the time evolution \cite{bloch:1934,pt:2013a}.

In the approach of operator-valued fields, used particularly in connection with the Wightman axioms \cite{WG64,RS2}, one employs the Heisenberg picture and thus regards the state vector $\psi_0$ as a fixed vector in some Hilbert space $\Hilbert_0$. The field operators $\Phi(x)$ act on $\Hilbert_0$ and are functions on space-time, $x=(t,\vx)\in\RRR^4$; all dynamics lies in the non-trivial $t$-dependence of $\Phi(t,\vx)$.

Our approach, in terms of multi-time wave functions $\phi(x_1,\ldots,x_N)$ with a variable $N$, can be regarded as a Schr\"odinger picture of QFT. We suggest in this paper that the three pictures are equivalent and provide several theoretical results in support of this suggestion. In particular, the three pictures can be translated into each other as follows. 

A multi-time wave function $\phi$ naturally defines a $\psi_\Sigma$ for every spacelike hypersurface $\Sigma$ by means of restriction to configurations consisting exclusively of points on $\Sigma$; in formulas, the function $\phi$ on the set $\sS$ of spacelike configurations defines the function $\psi_\Sigma$ on $\Gamma(\Sigma)$ by
\be\label{psiSigmadef}
\psi_\Sigma (x_1,\ldots,x_N) = \phi(x_1,\ldots,x_N) \quad \forall x_1,\ldots,x_N\in\Sigma.
\ee
The resulting $\psi_\Sigma$ belongs to a Hilbert space $\Hilbert_\Sigma$ containing functions on the configuration space $\conf_\Sigma =\Gamma(\Sigma)$ associated with $\Sigma$. Translating these $\psi_\Sigma$ into the interaction picture, we obtain the $\tilde\psi_\Sigma$ used in the Tomonaga--Schwinger approach. We show for the emission--absorption model that the $\tilde\psi_\Sigma$ obtained in this way from a multi-time wave function $\phi$ indeed satisfies the Tomonaga--Schwinger equation \eqref{TS}. G\"unther \cite{Gue52} reached an analogous conclusion for the hybrid model of Dirac, Fock, and Podolsky \cite{dfp:1932} described in Section~\ref{sec:literature} below.

Conversely, however, not every family $\{\psi_\Sigma\}_\Sigma$ of wave functions associated with hypersurfaces can be thought of as arising from a multi-time wave function $\phi$ via \eqref{psiSigmadef}. The $\psi_\Sigma$ defined through \eqref{psiSigmadef} has the property that, for any configuration $(x_1,\ldots,x_N)$ on $\Sigma$ and any other spacelike hypersurface $\Sigma'$ that also contains each of the points $x_1,\ldots,x_N$, 
\be\label{TStomultitime}
\psi_{\Sigma'}(x_1,\ldots,x_N) = \psi_\Sigma(x_1,\ldots,x_N).
\ee
If and only if a family $\{\psi_\Sigma\}_\Sigma$ has this property, then it can be converted into a multi-time wave function $\phi$, as then and only then it unambiguously defines $\phi(x_1,\ldots,x_N)$ via \eqref{psiSigmadef}.
We show that under certain assumptions on $\Hd_I(x)$, which we believe will often be satisfied, the Tomonaga--Schwinger equation implies \eqref{TStomultitime} and thus consistently defines a multi-time wave function $\phi$.

As noted essentially already by Bloch \cite{bloch:1934}, the physical relevance of the multi-time wave function $\phi$ (i.e., its connection with experiment) lies in that if we place detectors along the spacelike hypersurface $\Sigma$ then the probability density of which configuration in $\Gamma(\Sigma)$ to detect is given by $\rho=|\psi_\Sigma|^2$, suitably understood. (The exact expression for $\rho$ reads, in the example case of Dirac particles,
\be\label{rhophi}
\rho(x_1,\ldots,x_N)= \overline\phi(x_1,\ldots,x_N)\biggl( \prod_{i=1}^N n_\mu(x_i)\gamma_i^\mu \biggr) \phi(x_1,\ldots,x_N)\,,
\ee
where $x_1,\ldots,x_N\in\Sigma$, $n_\mu(x)$ is the future-pointing unit normal vector to $\Sigma$ at $x\in\Sigma$, $\gamma_i^\mu$ are the Dirac gamma matrices acting on the $i$-th spin index, and densities are expressed relative to the $3N$-volume on $\Sigma^N$ associated with the 3-metric on $\Sigma$. This expression equals $|\psi_\Sigma(x_1,\ldots,x_N)|^2$ if, for every $i$, the basis we use in the $i$-th spin space is the one corresponding to the Lorentz frame tangent to $\Sigma$ at $x_i$, i.e., whose spatial axes are tangent to $\Sigma$ at $x_i$.)

Concerning the approach of operator-valued fields in the Heisenberg picture, we suggest that the multi-time wave function $\phi$ is related to the operator-valued field $\Phi(x)$ on collision-free spacelike configurations according to
\be\label{phiPhi}
\phi(x_1,\ldots,x_N) = \scp{\emptyset}{\Phi(x_1)\cdots \Phi(x_N)|\psi_0}
\ee
if $\Phi$ is a bosonic field, with the understanding that $\Phi(x)=a(x) + a^\dagger(x)$ with $a$ the annihilation operator and $|\emptyset\rangle$ the vacuum state with the property $a(x)|\emptyset\rangle=0$. (In fact, if one uses ordered configurations, as we do, then a pre-factor $N!^{-1/2}$ occurs on the right-hand side of \eqref{phiPhi}, and similar pre-factors in \eqref{phiPhi2}--\eqref{phixyPhi}; see Section~\ref{sec:operatorvaluedfields}. Besides, we note that $\Phi$ can be replaced by $a$ in \eqref{phiPhi} because all terms involving $a^\dagger$ actually vanish.) Likewise,
\be\label{phiPhi2}
\phi\Bigl(x_1,\ldots,x_M,\overline{x}_1,\ldots,\overline{x}_N\Bigr)
= \Bscp{\emptyset}{\Phi(x_1)\cdots \Phi(x_M) \, \Phi^\dagger(\overline{x}_1) \cdots \Phi^\dagger(\overline{x}_N)\Big|\psi_0}
\ee
if $\Phi$ is a fermionic field, with the understanding that $\overline{x}_k$ is the coordinate of the $k$-th anti-particle, just as $x_j$ is that of the $j$-th particle, and that $\Phi(x)=a(x) + b^\dagger(x)$, with $a$ the particle annihilator and $b$ the anti-particle annihilator. Note that Equations~\eqref{phiPhi} and \eqref{phiPhi2} form natural generalizations of the following relation that always holds between a vector $|\psi\rangle$ in (bosonic or fermionic) Fock space and its particle--position representation on collision-free configurations:
\be\label{psiPhi}
\psi(\vx_1,\ldots,\vx_N) = \scp{\emptyset}{\Phi(\vx_1)\cdots \Phi(\vx_N)|\psi}
\ee
with field operators $\Phi(\vx)=a(\vx) + a^\dagger(\vx)$. Equation~\eqref{phiPhi} was first suggested, as far as we know, by Schweber \cite[p.~171]{schweber:1961}. If there are two particle species, say $x$-particles and $y$-particles with associated operator-valued fields $\Phi_x$ and $\Phi_y$, then
\be\label{phixyPhi}
\phi(x_1,\ldots,x_M,y_1,\ldots,y_N) = \Bscp{\emptyset}{\Phi_x(x_1)\cdots \Phi_x(x_M)\Phi_y(y_1)\cdots \Phi_y(y_N) \Big| \psi_0}\,,
\ee
if both fields are bosonic, and \emph{mutatis mutandis} if one of them is fermionic, or if more than two fields are involved. Here, we take for granted that $\Phi_x(x)$ commutes with $\Phi_y(y)$ 
whenever $x$ and $y$ are spacelike separated or $x=y$. 
(We also note that \eqref{phiPhi}, \eqref{phiPhi2}, and \eqref{phixyPhi} fit nicely with our earlier claim that multi-time wave functions $\phi$ are anti-symmetric (respectively, symmetric) against permutation of the space-time points of the particles belonging to a fermionic (respectively, bosonic) species; indeed, this follows from the fact that the field operators $\Phi(x_1),\Phi(x_2)$ anti-commute (respectively, commute) when $x_1,x_2$ are spacelike separated.)

In support of the scheme based on \eqref{phiPhi}, we show that \eqref{phixyPhi} is true in the emission--absorption model.\footnote{Although in this model, the $x$-particles are fermions, we ignore their antiparticles for the sake of simplicity; that is why \eqref{phixyPhi} applies without any $\Phi^\dagger(\overline{x}_j)$ factors.} That is, rather than taking \eqref{phixyPhi} as the definition of $\phi$, we start from the multi-time equations for the evolution of $\phi$ from initial datum $\psi_0$ and then find the result to be equal to the right-hand side of \eqref{phixyPhi}.

Several problems will be left aside in this paper. First, we leave aside the ultraviolet (UV) problem of QFT, i.e., the problem that the interaction Hamiltonian is usually mathematically ill-defined if electrons (and other particles) are regarded as point particles; we will mostly ignore the problem and proceed by formal calculation. Second, the solutions of the Dirac equation with negative energy are usually taken to be unphysical; for the purpose of examples of multi-time equations, we will make no effort to exclude wave functions with contributions of negative energy because this issue is orthogonal to those relevant to multi-time equations. Third, although the motivation for multi-time wave functions comes from relativity, we will consider examples of multi-time equations that are not (or not fully) relativistic; that is partly for simplicity and partly for the other difficulties mentioned.

\subsection{Comparison of Tomonaga--Schwinger Representation and Multi-Time Wave Functions}

As explained above, the Tomonaga--Schwinger picture and the framework of multi-time wave functions are very similar, and in most cases equivalent. At the same time, multi-time wave functions and their evolution equations are mathematically simpler than the wave functions and equations of the Tomonaga--Schwinger picture: First, the multi-time wave function is just a function of finitely many variables (more precisely, it is defined either on the space $\Gamma(\RRR^4)$ or on $\sS$, which have locally finite dimension) whereas $\tilde\psi_\Sigma$ depends on the spacelike hypersurface $\Sigma$, and the space of all spacelike hypersurfaces is infinite-dimensional. Second, the multi-time equations are just coupled partial differential equations, see \eqref{multi12} for example, whereas the Tomonaga--Schwinger equation \eqref{TS} is more sophisticated, involving the variation of a hypersurface in space-time.

On the other hand, multi-time equations have the following limitation that may sometimes be undesirable: They are linked to a particle representation of the quantum state. To illustrate this, let us consider a field representation of a quantum state; the most explicit way of doing this is the following: Consider a bosonic field, say a scalar field, classically described by a function $F:\RRR^4\to\RRR$. At any time $t$, the classical description of the field is given by a function $F:\RRR^3\to\RRR$; the quantum state could then be regarded as a function $\Psi(F)$; that is, as a function $\Psi:\conf\to\CCC$ on a suitable space $\conf$ of field configurations (i.e., of functions $\RRR^3\to\RRR$). Since there is one degree of freedom associated with each point of 3-space, a multi-time version of this representation should involve a choice of a point in time for every point in 3-space---a function $t(\vx)$ representing a hypersurface in space-time. (This proposal was, in fact, made by Dirac, Fock, and Podolsky \cite{dfp:1932}.) That is why the Tomonaga--Schwinger picture arises from a field representation in much the same way as the multi-time framework as in \eqref{phi} from a particle representation. Yet, the Tomonaga--Schwinger picture is so flexible that one can also consider a particle representation of $\psi_\Sigma$, regarding $\psi_\Sigma$ (as we did above already) as a function on $\Gamma(\Sigma)$---or even, for a fixed number $N$ of particles, as a function on $\Sigma^N$.

Finally, there is another respect in which the multi-time wave function $\phi$ works better than the Tomonaga--Schwinger wave function $\psi_\Sigma$ (or $\tilde\psi_\Sigma$): it works with a UV cut-off. Such a cut-off usually amounts to giving electrons (and other particles) a positive radius $\delta$ and smearing out the interaction Hamiltonian over this radius in some preferred Lorentz frame. We describe in \cite{pt:2013b} how to set up multi-time equations with UV cut-off. The Tomonaga--Schwinger picture, in contrast, does not work with a UV cut-off; that is because the cut-off interaction Hamiltonian of an electron at $x$ acts on a 3-dimensional $\delta$-neighborhood of $x$ at the same time in the preferred frame, and this 3-d set is not contained in most hypersurfaces $\Sigma$ containing $x$. Tomonaga was well aware of this:
\begin{quotation}
[W]hat bearing would have the so-called cut-off hypothesis on our theory? [...] Miyazima has once noticed that, although our theory seems at first sight to allow the introduction of a relativistically invariant cut-off process, taking as $H_{I,II}(P)$ not the energy density just at the world point $P$ but some average over a finite world region surrounding $P$, such a procedure breaks necessarily the condition of integrability of the fundamental equation (III). \cite[p.~207]{KTT:1947b}
\end{quotation}

\subsection{Prior Works}
\label{sec:literature}

Multi-time wave functions were considered early on in the history of quantum theory \cite{dirac:1932,dfp:1932,bloch:1934,Gue52} but apparently have never been studied comprehensively. Multi-time wave functions were first suggested by Dirac \cite{dirac:1932} in 1932 and taken up by Dirac, Fock, and Podolsky \cite{dfp:1932} in the same year. However, in these papers there is no mention that multi-time equations are inconsistent unless the consistency conditions are satisfied. As noted above, they used a hybrid of a field representation and a particle representation: the quantized radiation field was expressed in a field representation and the electrons in a particle representation, leading to a wave function of the form
\be\label{DFP}
\Psi\Bigl(x_1,\ldots,x_N, (F(x))_{x\in\Sigma} \Bigr)\,,
\ee
i.e., a function of $N$ space-time points for the electrons and of a field configuration on a spacelike hypersurface $\Sigma$.

Bloch \cite{bloch:1934} was the first to discuss consistency conditions, to suggest that the multi-time wave function $\phi$ should be defined only on the spacelike configurations, and to describe the relation between $\phi$ and the probability distribution of the outcomes of experiments. G\"unther \cite{Gue52} further studied the hybrid model of Dirac, Fock, and Podolsky; he verified the equivalence to the appropriate Tomonaga--Schwinger equation and calculated applications to bound states. Schweber \cite[p.~171]{schweber:1961} may have been the first to consider a wave function with a variable number of time variables, in fact a wave function on $\Gamma(\RRR^4)$. Marx \cite{Marx:1972} thought about the permutation symmetry of multi-time wave functions and drew the (somewhat too radical) conclusion that one needs a modified Fock space of asymmetric wave functions; see also our discussion in Section~\ref{sec:permutation}. Droz-Vincent \cite{DV82b,DV85} also considered a wave function with a variable number of time variables in the particle--position representation (on $\Gamma(\RRR^4)$, though not on the set $\sS$ of spacelike configurations). He provided an example of a consistent multi-time evolution on $\Gamma(\RRR^4)$ with interaction, which however does not correspond to any known QFT and cannot be written down as a system of partial differential equations. In fact, the evolution is non-local in time (i.e., involves integration of time variables over the entire real axis). As a consequence, it cannot be formulated as an initial-value problem; i.e., the wave function is not determined by its values (or its time derivatives up to some order $n$) at time 0 (i.e., at all times 0). Nikoli\'c~\cite{Nik10} suggested to take \eqref{phiPhi}, with $\Phi$ replaced by $a$, as the definition of a multi-time wave function also for non-spacelike configurations.

See \cite{pt:2013a} for no-go results about interaction potentials and for references to works about multi-time wave functions outside of QFT. See \cite{pt:2013e} for a comparison of the status and significance of multi-time formulations in classical and quantum physics.

\bigskip

The remainder of this article is organized as follows. In Section~\ref{sec:modelQFT}, we develop a consistent set of multi-time equations for a model QFT involving particle creation and annihilation (the emission--absorption model). In Section~\ref{sec:operatorvaluedfields}, we study the connection between multi-time wave functions and operator-valued fields. In Section~\ref{sec:TS}, we study the connection between our multi-time equations and the Tomonaga--Schwinger equation. Detailed calculations are postponed to Section~\ref{sec:derivations}. Our presentation includes a number of theorems and proofs, which however are not mathematically rigorous and therefore called assertions and derivations.

\section{Emission--Absorption Model: A Model QFT}
\label{sec:modelQFT}

The emission--absorption model is adapted from \cite[p.~339]{schweber:1961} and \cite{Nel64}. It involves two species of particles, $x$ and $y$; the $x$-particles can emit and absorb $y$-particles. For simplicity, we take both species to be Dirac particles. 

\subsection{One-Time Formulation}

The Hilbert space is a tensor product of two Fock spaces,
\be\label{Hilbertdef}
\Hilbert = \Hilbert_x \otimes \Hilbert_y
\ee
with
\be\label{Fockdef}
\Hilbert_{x,y} = \bigoplus_{N=0}^\infty S_{x,y} L^2(\RRR^3,\CCC^4)^{\otimes N}
\ee
(where the subscript $x,y$ means either $x$ or $y$). Here, $S_x$ is (say) the anti-symmetrization operator and $S_y$ the symmetrization operator (so $x$-particles are fermions and $y$-particles are bosons\footnote{This choice is contrary to the spin--statistics relation; but that does not matter for the purposes of the emission--absorption model. We choose spin $\frac12$ for the $y$-particles to avoid other problems that arise for photon wave functions.}); thus, $S_y L^2(\RRR^3,\CCC^4)^{\otimes N}$ is the range of the symmetrization operator, i.e., the space of symmetric elements in $L^2(\RRR^3,\CCC^4)^{\otimes N}$. A vector in $\Hilbert$ can be regarded as a function $\psi$ on the configuration space $\conf=\conf_x\times \conf_y$ with $\conf_x=\conf_y=\Gamma(\RRR^3)$, the \emph{particle-position representation}. The function $\psi$ can be thought of as a cross-section of the vector bundle with fiber $(\CCC^4)^{\otimes M} \otimes (\CCC^4)^{\otimes N}$ over the sector $(\RRR^3)^M\times (\RRR^3)^N$ of configuration space $\conf=\Gamma(\RRR^3)\times\Gamma(\RRR^3)=\Gamma(\RRR^3)^2$. A generic element of $\Gamma(\RRR^3)^2$ can be written as $q^3=(x^{3M},y^{3N})=(\vx_1,\ldots, \vx_M, \vy_1,\ldots,\vy_N)$, where bold-face symbols denote 3-vectors, while $x^{3M}$ denotes a configuration of $M$ $x$-particles in $\RRR^3$ and $y^{3N}$ one of $N$ $y$-particles. We also write $\psi_t(x^{3M},y^{3N})$ for $\psi(t,x^{3M},y^{3N})$, or, when we want to make the spin indices explicit, $\psi_{r_1\ldots r_M,s_1\ldots s_N}(x^{3M},y^{3N})$ for $\psi(x^{3M},y^{3N})$; sometimes we want to make just one of the spin indices explicit and write (e.g.) $\psi_{s_N}(x^{3M},y^{3N})$ for the same object.

In the single-time version of the model, $\psi$ evolves according to the Schr\"odinger equation
\begin{subequations}\label{Schr1Hdef}
\be\label{Schr1}
i\frac{\partial}{\partial t}\psi_t(x^{3M},y^{3N}) = (H\psi_t)(x^{3M},y^{3N})
\ee
with Hamiltonian $H$ given by
\begin{align}
(H\psi)(x^{3M},y^{3N}) 
&= \sum_{j=1}^M H^\free_{x_j} \psi(x^{3M},y^{3N}) + \sum_{k=1}^N H^\free_{y_k} \psi(x^{3M},y^{3N})\nonumber\\
&\quad + \sqrt{N+1} \sum_{j=1}^M \sum_{s_{N+1}=1}^4 g_{s_{N+1}}^* \, \psi_{s_{N+1}}\bigl(x^{3M}, (y^{3N}, \vx_j)\bigr)\nonumber \\
& \quad + \frac{1}{\sqrt{N}} \sum_{j=1}^M \sum_{k=1}^N g_{s_k}\, \delta^3(\vy_k - \vx_j) \, \psi_{\widehat{s_k}}(x^{3M}, y^{3N} \backslash \vy_k) \,.
\label{Hdef}
\end{align}
\end{subequations}
Here, $\widehat{s_k}$ means that the index $s_k$ is omitted, $g\in\CCC^4$ is a given spinor, and we use the notation 
\be\label{ysetminusdef}
y^{3N}\setminus \vy_k = (\vy_1,\ldots,\vy_{k-1},\vy_{k+1},\ldots,\vy_N)
\ee
for the configuration of $N-1$ $y$-particles with the $k$-th particle removed. Note that the expression $\psi(x^{3M},y^{3N}\setminus \vy_k)$ refers to the sector of $\psi$ with $N-1$ $y$-particles (sometimes denoted $\psi^{(M,N-1)}$), and $\psi\bigl(x^{3M},(y^{3N},\vx_j)\bigr)$ to the sector of $\psi$ with $N+1$ $y$-particles; thus, the time evolution couples different sectors of $\psi$. For $N=0$, the last line of \eqref{Hdef}, involving a sum of 0 terms, is understood to be 0.
The free Hamiltonians are Dirac operators, 
\begin{align}
H^\free_{x_j} \psi_{r_j}(x^{3M},y^{3N}) &= \sum_{r_j'=1}^4\biggl( -i \sum_{a=1}^3  (\alpha_a)_{r_jr_j'} \, \frac{\partial}{\partial x_j^a} + m_x  \beta_{r_jr_j'} \biggr) \psi_{r_j'}(x^{3M},y^{3N})\label{Hxjfreedef} \\
H^\free_{y_k} \psi_{s_k}(x^{3M},y^{3N}) &= \sum_{s_k'=1}^4\biggl( -i \sum_{a=1}^3  (\alpha_a)_{s_ks_k'} \, \frac{\partial}{\partial y_k^a} + m_y  \beta_{s_ks_k'} \biggr) \psi_{s_k'}(x^{3M},y^{3N})\label{Hykfreedef}
\end{align}
with mass parameters $m_x,m_y \geq 0$.

The Hamiltonian is of the form $H=H_x+H_y+H_\inter$, where $H_x\psi(x^{3M},y^{3N})=\sum_{j=1}^M H_{x_j}^\free \psi(x^{3M},y^{3N})$ and $H_y\psi(x^{3M},y^{3N})=\sum_{k=1}^N H^\free_{y_k} \psi(x^{3M},y^{3N})$ are the free Hamiltonians, and the second and third line of \eqref{Hdef} form the interaction Hamiltonian responsible for the creation and annihilation of $y$-particles. In terms of creation and annihilation operators, and with $\valpha=(\alpha_1,\alpha_2,\alpha_3)$,
\begin{align}
H_{x} &= \int d^3\vx \sum_{r,r'=1}^4 a_r^\dagger(\vx) \bigl(-i\valpha_{rr'}\cdot \nabla+m_x\beta_{rr'}\bigr) a_{r'}(\vx)\\
H_{y} &= \int d^3\vx \sum_{s,s'=1}^4 b_s^\dagger(\vx) \bigl(-i\valpha_{ss'}\cdot\nabla+m_y \beta_{ss'}\bigr) b_{s'}(\vx) \\
H_\inter &=  \int d^3\vx\sum_{r,s=1}^4 a_r^\dagger(\vx) \, \bigl(g_s^*\,b_s(\vx)+g_s\,b_s^\dagger(\vx) \bigr)\, a_r(\vx)
\end{align}
with $^\dagger$ denoting the adjoint operator, and $a_s(\vx), b_s(\vx)$ the annihilation operators for an $x,y$-particle with spin component $s$ at location $\vx$ in position space, explicitly defined by
\begin{align}
\bigl(a_r(\vx)\,\psi\bigr)(x^{3M},y^{3N}) 
&= \sqrt{M+1}\; (-1)^M\, \psi_{r_{M+1}=r}\bigl((x^{3M},\vx),y^{3N}\bigr)\label{adef}\\
\bigl(a_r^\dagger(\vx)\,\psi\bigr)(x^{3M},y^{3N}) 
&= \frac{1}{\sqrt{M}} \sum_{j=1}^M (-1)^{j+1} \, \delta_{rr_j}\,\delta^3(\vx_j-\vx)\, \psi_{\widehat{r_j}}\bigl(x^{3M}\setminus \vx_j,y^{3N}\bigr)\,,\\
\bigl(b_s(\vx)\,\psi\bigr)(x^{3M},y^{3N}) 
&= \sqrt{N+1}\; \psi_{s_{N+1}=s}\bigl(x^{3M},(y^{3N},\vx)\bigr)\label{bdef}\\
\bigl(b_s^\dagger(\vx)\,\psi\bigr)(x^{3M},y^{3N}) 
&= \frac{1}{\sqrt{N}} \sum_{k=1}^N \delta_{ss_k}\, \delta^3(\vy_k-\vx)\,\psi_{\widehat{s_k}}\bigl(x^{3M},y^{3N}\setminus \vy_k\bigr)\,.
\end{align}
(The combinatorial factors $\sqrt{M+1}$, $M^{-1/2}$, etc.\ arise from the fact that every configuration of $M$ $x$-particles occurs in $M!$ different permutations. They would be absent if we used unordered configurations \cite[section 2.4]{GTTZ:2013}.)

\bigskip

\noindent\textbf{Remarks.}
\begin{enumerate}
\setcounter{enumi}{\theremarks}
\item The emission--absorption model is not (fully) Lorentz invariant, for example because the spinor $g$ will transform to a different spinor in another Lorentz frame; put differently, every choice of a fixed spinor $g\in\CCC^4$ selects a preferred frame. 
\item We ignore the fact that the Hamiltonian is ultraviolet-divergent and therefore mathematically ill defined. Also, for the sake of simplicity we do not exclude states of negative energy from Hilbert space; we leave anti-particles aside.
\end{enumerate}
\setcounter{remarks}{\theenumi}

\subsection{Multi-Time Formulation}
\label{sec:MT-em-ab}

We write $x^{4M}=(x_1,\ldots,x_M)$ and $y^{4N}=(y_1,\ldots,y_N)$ for configurations of space-time points $x_j,y_k\in\RRR^4$.
For two particle species, a spacelike configuration is given by an element of $\sS$ (i.e., any number of spacelike-separated points, possibly with repetitions) with each point marked as either an $x$-particle or a $y$-particle; we denote the set of spacelike two-species configurations by $\sS_{xy}$. Equivalently, with a slight abuse of notation (viz., assuming that every point called $x_j$ is marked as an $x$-particle and every point called $y_k$ as a $y$-particle), we can define $\sS_{xy}$ by
\begin{align}
\sS_{xy} = \bigcup_{M,N=0}^\infty \Bigl\{ 
(x_1,&\ldots,x_M,\, y_1,\ldots,y_N)\in(\RRR^4)^{M}\times (\RRR^4)^N: \nonumber\\
&(1)\; \forall i\neq j\in\{1\ldots M\}: x_i \sim x_j \text{ or } x_i=x_j\nonumber\\[4mm]
&(2)\; \forall k\neq \ell\in\{1\ldots N\}: y_k \sim y_\ell \text{ or } y_k = y_\ell\nonumber\\[2mm]
&(3)\; \forall j\in\{1\ldots M\} \forall k\in\{1\ldots N\}: x_j\sim y_k \text{ or } x_j=y_k\Bigr\}\,.
\label{sSxydef}
\end{align}
In this way, we can regard $\sS_{xy}$ as a subset of $\Gamma(\RRR^4)\times \Gamma(\RRR^4)=\Gamma(\RRR^4)^2$ and write $q^4=(x^{4M},y^{4N})$ for $q^4\in\sS_{xy}$. 

The multi-time version of the emission--absorption model that we propose is defined as follows.
The desired multi-time wave function $\phi$ is a function on $\sS_{xy}$ 
with values $\phi(x^{4M},y^{4N})\in (\CCC^4)^{\otimes M} \otimes (\CCC^4)^{\otimes N}$. The wave function $\phi$ obeys the following multi-time equations (again with a given spinor $g\in\CCC^4$):
\begin{subequations}\label{multi12}
\begin{align}\label{multi1}
i \frac{\partial\phi}{\partial x_j^0}(x^{4M}, y^{4N}) &= H^\free_{x_j} \phi(x^{4M}, y^{4N}) 
+ \sqrt{N+1}  \sum_{s_{N+1}=1}^4 g_{s_{N+1}}^* \, \phi_{s_{N+1}}\bigl(x^{4M}, (y^{4N}, x_j)\bigr)\nonumber \\
& \quad + \frac{1}{\sqrt{N}} \sum_{k=1}^N G_{s_k}(y_k - x_j) \, \phi_{\widehat{s_k}}\bigl(x^{4M}, y^{4N} \backslash y_k\bigr)  \\[3mm] 
\label{multi2}
i \frac{\partial\phi}{\partial y_k^0}(x^{4M}, y^{4N}) &= H^\free_{y_k}\phi(x^{4M}, y^{4N}),
\end{align}
\end{subequations}
where $G:\RRR^4\to\CCC^4$ is a Green function, i.e., the solution of
\be\label{Gdef1}
i\frac{\partial G}{\partial t} = H^\free_y \, G
\ee
with initial condition
\be\label{Gdef2}
G_s(0,\vy) = g_s \delta^3(\vy)\,. 
\ee
(The function $G(y-x_j)$ can be thought of as the wave function of a $y$-particle created at $x_j$; this view makes it plausible that $G$ should satisfy the evolution equation of a single $y$-particle, and should start out as a delta function.) 

For $N=0$, the last line in \eqref{multi1}, involving a sum of 0 terms, is understood to be 0.

\begin{ass}\label{ass:consistency}
On a non-rigorous level ignoring the ultraviolet divergence,
the multi-time system \eqref{multi12} is consistent on $\sS_{xy}$, i.e., it possesses a unique solution $\phi$ on $\sS_{xy}$ for every given initial datum $\phi_0$ (on $\Gamma(\RRR^3)^2$, i.e., setting all times to zero). If $g^\dagger \beta g =0$ (i.e., $\sum_{s,s'}g^*_s \beta_{ss'} g_{s'}=0$) or $m_y=0$, then the system is also consistent on non-spacelike configurations, i.e., the solutions can be extended to all of $\Gamma(\RRR^4)^2$, but not so if $g^\dagger \beta g\neq 0$ and $m_y>0$.
\end{ass}

The derivation is given in Section~\ref{sec:derivations}.

\bigskip

\noindent\textbf{Remarks.}
\begin{enumerate}
\setcounter{enumi}{\theremarks}
\item While it so happens that this particular multi-time system \eqref{multi12} is consistent also on \emph{non}-spacelike configurations for special choices of $g\in\CCC^4$ or $m_y=0$, we expect that generically, multi-time equations will be consistent only on spacelike configurations. 
\item\label{rem:consistency} The consistency is closely related to the condition
\be\label{xxconsistency1}
\biggl[ i\frac{\partial}{\partial x_i^0} -H_{x_i}, i\frac{\partial}{\partial x_j^0} - H_{x_j} \biggr] =0
\ee
for $i\neq j$ and the corresponding sister conditions between $y_k$ and $y_\ell$ ($k\neq \ell$) and between $x_j$ and $y_k$. It is known \cite{pt:2013a} that for a fixed number $N$ of particles, the analogous condition is necessary and sufficient for the consistency of the multi-time equations. As explained in Section~\ref{sec:non-spacelike}, the fact that $\partial^2 \phi/\partial x_i^0 \partial x_j^0 = \partial^2 \phi/\partial x_j^0 \partial x_i^0$ yields that \eqref{xxconsistency1} is a necessary condition. Part of the derivation of Assertion~\ref{ass:consistency} is to compute the commutators in \eqref{xxconsistency1} and the sister conditions and verify that they always vanish on spacelike configurations,\footnote{\label{fn:non-interact}That they vanish is intuitively plausible on collision-free spacelike configurations, as there the particles do not interact, but less obvious at collision configurations.} but elsewhere only if $m_y=0$ or $g^\dagger \beta g=0$. Another part of the derivation is devoted to showing that \eqref{xxconsistency1} and the sister conditions are also \emph{sufficient} for consistency. Since at any collision-free spacelike configuration $q^4$, as mentioned in Footnote~\ref{fn:non-interact}, the particles are non-interacting, it is intuitively plausible that \eqref{multi12} is consistent in a neighborhood of $q^4$. However, consistency on all of $\sS_{xy}$ is rather non-trivial, for several reasons: consistency near a collision configuration is less obvious; the number of time variables is not fixed; we typically have \eqref{xxconsistency1} only on \emph{spacelike} configurations; and we need to solve \eqref{multi12} in such an ``ordering'' of configurations that the terms referring to other sectors of $\phi$ are always uniquely defined when they are used.

\item\label{rem:tip} We need to clarify the meaning of the equations \eqref{multi12} at the tips of $\sS_{xy}$, i.e., at collision configurations. That is, while the derivatives $\partial \phi/\partial x_j^0$ and $\partial \phi/\partial y_k^0$ make immediate sense at interior points of $\sS_{xy}$ (i.e., collision-free spacelike configurations), they do not at collision configurations. Suppose, for example, that $x_j=y_k$. Since the derivative $\partial/\partial x_j^0$ refers to varying $x_j^0$ while keeping all other variables fixed, and since this varying will lead off of $\sS_{xy}$, $\partial \phi/\partial x_j^0$ is not defined at such a configuration. However, if we change $x_j$ and $y_k$ by the same amount $\delta x$, then the configuration stays in $\sS_{xy}$ (at least when all other particles are spacelike separated), so $(\partial/\partial x_j^0 + \partial/\partial y_k^0)\phi$ exists in the sense of a directional derivative. The natural understanding of the multi-time equations \eqref{multi12}, which we will henceforth assume, is that all directional derivatives tangent to $\sS_{xy}$ have the value corresponding to the appropriate linear combination of \eqref{multi12}. Thus, at a configuration with $x_j=y_k$, the Equations~\eqref{multi12} mean that 
\begin{align}
i \biggl(\frac{\partial}{\partial x_j^0}+\frac{\partial}{\partial y_k^0}\biggr) \phi(x^{4M}, y^{4N}) 
&= H^\free_{x_j} \phi(x^{4M}, y^{4N}) + H^\free_{y_k}\phi(x^{4M}, y^{4N}) \nonumber\\
&\quad + \sqrt{N+1}  \sum_{s_{N+1}=1}^4 g_{s_{N+1}}^* \, \phi_{s_{N+1}}\bigl(x^{4M}, (y^{4N}, x_j)\bigr)\nonumber \\
& \quad + \frac{1}{\sqrt{N}} \sum_{k=1}^N G_{s_k}(y_k - x_j) \, \phi_{\widehat{s_k}}(x^{4M}, y^{4N} \backslash y_k)  \,.
\label{multi5}
\end{align}
\item If we set all time variables equal in $\phi$,
\be\label{psitphi}
\psi(t,\vx_1,\ldots,\vx_M,\vy_1,\ldots,\vy_N) = \phi(t,\vx_1,\ldots,t,\vx_M,t,\vy_1,\ldots,t,\vy_N)\,,
\ee
the 1-time wave function $\psi$ obtained is the usual 1-time wave function. Indeed, $\psi_t$ coincides with the wave function $\psi_\Sigma$ associated with the hypersurface $\Sigma=\{t=\mathrm{const.}\}$ as in \eqref{psiSigmadef}. Note further that, as a consequence of \eqref{psitphi} and \eqref{multi12}, $\psi$ evolves according to the 1-time Schr\"odinger equation \eqref{Schr1} with Hamiltonian \eqref{Hdef}. That is because the right-hand sides of \eqref{multi12} add up, when all times are equal, to $H\psi$. (But see also Remark~\ref{rem:0particles}.) For this reason, we also write $H_{x_j}\phi$ for the right-hand side of \eqref{multi1} and $H_{y_k}\phi$ for that of \eqref{multi2}, and call $H_{x_j}$ and $H_{y_k}$ the \emph{partial Hamiltonians}. 

Conversely, when setting up a multi-time model, the wish that it contain a particular 1-time theory (e.g., the one given by \eqref{Schr1Hdef}) leads to the demand that the partial Hamiltonians are chosen so that they add up to $H$, thus providing a method for guessing multi-time equations: For each term in $H$, decide to which particle it belongs. In \eqref{Hdef}, this is obvious for the free Hamiltonians; the second line of \eqref{Hdef} is a sum of $M$ terms, the $j$-th of which naturally belongs to $x_j$; the last line of \eqref{Hdef} is more ambiguous, as it is a sum of $MN$ terms, each of which belongs to some $x_j$ and $y_k$. One possibility is to attribute to $x_j$ the sum over all $y_k$; this leads to \eqref{multi12}. Another, equally natural, possibility is to attribute to $y_k$ the sum over all $x_j$; this leads to, instead of \eqref{multi12},
\begin{subequations}\label{multi67}
\begin{align}\label{multi6}
i \frac{\partial\phi}{\partial x_j^0}(x^{4M}, y^{4N}) &= H^\free_{x_j} \phi(x^{4M}, y^{4N}) 
+ \sqrt{N+1}  \sum_{s_{N+1}=1}^4 g_{s_{N+1}}^* \, \phi_{s_{N+1}}\bigl(x^{4M}, (y^{4N}, x_j)\bigr)\\
\label{multi7}
i \frac{\partial\phi}{\partial y_k^0}(x^{4M}, y^{4N}) &= H^\free_{y_k}\phi(x^{4M}, y^{4N})\nonumber\\
& \quad + \frac{1}{\sqrt{N}} \sum_{j=1}^M G_{s_k}(y_k - x_j) \, \phi_{\widehat{s_k}}(x^{4M}, y^{4N} \backslash y_k)  \,.
\end{align}
\end{subequations}
These equations differ from \eqref{multi12} only in that one term has been removed from the $x_j$-equation and added instead (suitably modified) to the $y_k$-equation, and that this term now involves a sum over $j$ instead of $k$.

Appearances to the contrary notwithstanding, the system of equations \eqref{multi67} is actually equivalent to \eqref{multi12} on $\sS_{xy}$. To see this, note first that the Green function $G$ vanishes whenever $y_k\sim x_j$, so that the terms by which \eqref{multi67} differ from \eqref{multi12} vanish in the interior of $\sS_{xy}$ (i.e., on collision-free configurations). At the tip corresponding to $x_j=y_k$, however, with the understanding of directional derivatives discussed in Remark~\ref{rem:tip}, the system \eqref{multi67} provides the same prescription \eqref{multi5} as the system \eqref{multi12}.\footnote{Nevertheless, the system \eqref{multi67} is \emph{not} equivalent to \eqref{multi12} on $\Gamma(\RRR^4)^2$ (which includes \emph{non}-spacelike configurations). Indeed, whenever both systems are consistent on the latter set, the derivative (e.g.)\ $\partial \phi/\partial y_k^0$ can be computed everywhere without restrictions, so it must be different for solutions of \eqref{multi7} than of \eqref{multi2}.}

Another way of seeing that \eqref{multi67} is equivalent to \eqref{multi12} on $\sS_{xy}$ is that both are equivalent to the same Tomonaga--Schwinger equation, see Section~\ref{sec:TSfromMT}.

\item Unlike $H$, the partial Hamiltonians are not operators on Hilbert space in the usual sense. That is because $H_{x_j}\phi$ involves inserting $x_j$ as the time and place of the $N+1$-st $y$-particle into the $(M,N+1)$-sector of $\phi$. The Hilbert space contains functions of the space variables, while the time variable is (or time variables are) kept fixed. Therefore, one cannot insert a time coordinate (such as $x_j^0$) into a function from Hilbert space. 

Given that the $H_{x_j}$ and $H_{y_k}$ are not even operators on Hilbert space, one cannot ask whether they are self-adjoint. Still, $H^\free_{x_j}$ can of course be naturally thought of as a self-adjoint operator on Hilbert space. In contrast, the right-hand side of \eqref{multi6} does not even look self-adjoint, as the last term is a variant of the second line in \eqref{Hdef}, whose adjoint is the last line in \eqref{Hdef}, and whose counterpart in \eqref{multi67} is not part of $H_{x_j}$ but of $H_{y_k}$. 

\item\label{rem:0particles} There is a potential problem about turning a given 1-time theory into a multi-time theory: Consider the 0-particle sector (which means, in the emission--absorption model, the 00-particle sector) of $\psi$ and $\phi$. Since every sector of $\phi$ has as many time variables as particles, the 0-particle sector of $\phi$ is necessarily time-independent. However, since $\psi$ in its entirety is $t$-dependent, also its 0-particle sector can in principle be $t$-dependent. In this case, as a consequence, it is impossible to set up any multi-time equations such that $\psi$ is recovered from $\phi$ by setting all time variables equal. The problem does not arise if the 0-particle sector of $\psi$ is constant as a function of $t$, which occurs only if every vacuum state (i.e., one that vanishes outside the 0-particle sector) is an eigenstate of $H$ with eigenvalue 0. This is actually the case for $H$ as in \eqref{Hdef}. More generally (and leaving aside the trivial case that the vacuum is an eigenstate with eigenvalue $\neq 0$, which can be fixed by adding a constant to the Hamiltonian), this is the case if particles cannot be created out of the vacuum (and not annihilated into vacuum). The latter condition will often be violated in effective, phenomenological models, but should be obeyed in fundamental physical theories.
\item\label{rem:psit} In the 1-time theory, one often considers a vector in Hilbert space---a wave function of spatial variables only, as obtained from $\psi$ by inserting a particular value of $t$; in short, $\psi_t(\cdot)$. In the context of a multi-time wave function $\phi$, a natural analog is to specify $\phi$ on a spacelike hypersurface $\Sigma$, i.e., to specify $\psi_\Sigma(\cdot)$. Another analog that one might be inclined to consider \cite{DV82b,DV85} is obtained by inserting particular values for all time variables into $\phi$; that is, choose, for every $M,N\geq 0$ and every $j\in\{1,\ldots,M\}$, $k\in\{1,\ldots,N\}$, time values $X_{M,N,j}^0$ and $Y_{M,N,k}^0$, let $T$ denote the list of these values, and let $\phi_T$ be the function on $\Gamma(\RRR^3)^2$ obtained by inserting these values,
\be\label{phiT}
\phi_T(x^{3M},y^{3N})= 
\phi\bigl(X_{M,N,1}^0,\vx_1,\ldots,X_{M,N,M}^0,\vx_M,Y_{M,N,1}^0,\vy_1,\ldots,Y_{M,N,N}^0,\vy_N\bigr)\,.
\ee
However, the analogy is limited, and in fact not all that natural. 

First, not all combinations of time and spatial variables constitute spacelike configurations; thus, if $\phi$ is defined on $\sS_{xy}$, then $\phi_T$ is typically not defined on all of $\Gamma(\RRR^3)^2$, but entire regions of the spatial variables are excluded. Without them, insufficient information about $\phi$ is provided to determine $\phi$ on $\sS_{xy}$ (i.e., $\phi_T$ will not suffice as initial datum). In contrast, $\psi_\Sigma$ involves only spacelike configurations, determines $\phi$ on $\sS_{xy}$, and thus can serve as initial datum, see Assertion~\ref{ass:unitary_HS_evolution} in Section~\ref{sec:TS} below. 

Second, even in case $m_y=0$ or $g^\dagger \beta g=0$, when $\phi$ is defined not only on $\sS_{xy}$ but on all of $\Gamma(\RRR^4)^2$, and $\phi_T$ is defined on all of $\Gamma(\RRR^3)^2$, presumably $\phi_T$ still cannot serve as initial datum because it typically does not determine $\phi$. Indeed, to obtain $\phi(x^{4M},y^{4N})$ with $x_j^0$ in a neighborhood of $X_{M,N,j}^0$, we need to solve \eqref{multi1}, which requires, for computing $\partial\phi/\partial x_j^0$, the knowledge of $\phi\bigl(x^{4M},(y^{4N},x_j)\bigr)$ and $\phi(x^{4M},y^{4N}\setminus y_k)$, which may not be provided by $\phi_T$. (Of course, $\phi_T$ does determine $\phi$ for special choices of $T$ such as, with all time values equal.) As a consequence, while there is an evolution operator $U_{t\to t'}$ that maps $\psi_t$ to $\psi_{t'}$ (and one that maps $\psi_{\Sigma}$ to $\psi_{\Sigma'}$, see Assertion~\ref{ass:unitary_HS_evolution}), presumably there exists no operator that maps $\phi_T$ to $\phi_{T'}$ for generic choices $T,T'$ of the time values. Moreover, $\phi_T$ will typically not have $L^2$ norm 1 if $\phi_0$ has (again in contrast to $\psi_t$ and $\psi_\Sigma$, which are normalized). Indeed, when we keep all time values in $T$ fixed except for $X_{M,N,j}^0$, then all sectors of $\phi_T$ remain the same except for the $(M,N)$-sector, whose $L^2$ norm may well grow or shrink in response to the creation and annihilation terms in \eqref{multi1}. Note also that the lack of normalization of $\phi_T$ fits the situation that we have a version \eqref{rhophi} of Born's rule involving $\psi_\Sigma$ but none involving $\phi_T$.

\item\label{rem:g3} A natural generalization of the emission--absorption model is to replace the spinor $g_s$ by an object with 3 indices, $g_{srr'}$, and have it act on the spin index $r_j$ of the $x$-particle emitting or absorbing the $y$-particles. That is, in \eqref{multi1} replace in the first line
\begin{align}
\label{g3changes1}
\sum_{s_{N+1}=1}^4 g_{s_{N+1}}^* \, \phi_{s_{N+1}}
&\quad\longrightarrow\quad
\sum_{s_{N+1},r'_j=1}^4 g_{s_{N+1}r'_jr_j}^* \, \phi_{r'_js_{N+1}}\\
\intertext{and in the second line}
\label{g3changes2}
G_{s_k}(y_k - x_j) \, \phi_{\widehat{s_k}}
&\quad\longrightarrow\quad
\sum_{r'_j=1}^4 G_{s_kr_jr'_j}(y_k - x_j) \, \phi_{r'_j\widehat{s_k}}\,,
\end{align}
where $G:\RRR^4\to\CCC^4\otimes \CCC^{4\times 4}$ is the solution of
\be\label{Gdef1g3}
i\frac{\partial G_{srr'}}{\partial t} = \sum_{s'=1}^4 \biggl( -i \sum_{a=1}^3  (\alpha_a)_{ss'} \, \frac{\partial}{\partial y^a} + m_y  \beta_{ss'} \biggr) \, G_{s'rr'}(t,\vy)
\ee
with initial condition
\be\label{Gdef2g3}
G_{srr'}(0,\vy) = g_{srr'} \delta^3(\vy)\,. 
\ee
Leave \eqref{multi2} unchanged. The corresponding change of the single-time Hamiltonian \eqref{Hdef} can be obtained by setting all times equal. One can show in the same way as for Assertion~\ref{ass:consistency} that this modified system is always consistent on $\sS_{xy}$. 
\end{enumerate}
\setcounter{remarks}{\theenumi}

\subsection{Permutation Symmetry}
\label{sec:permutation}

\begin{ass}\label{ass:permutation}
The solutions $\phi$ of the multi-time equations \eqref{multi12} of the emission--absorption model  have the following permutation symmetry: For any integers $M,N\geq 0$ and any permutations $\pi$ of $\{1,\ldots,M\}$ and $\rho$ of $\{1,\ldots,N\}$, and with $(-1)^\pi$ denoting the sign of $\pi$,
\begin{multline}\label{permutation}
\phi_{r_{\pi(1)}\ldots r_{\pi(M)},s_{\rho(1)}\ldots s_{\rho(N)}}
\bigl(x_{\pi(1)},\ldots,x_{\pi(M)},y_{\rho(1)},\ldots,y_{\rho(N)}\bigr) =\\
 (-1)^\pi \phi_{r_1\ldots r_M,s_1\ldots s_N}\bigl(x_1,\ldots,x_M,y_1,\ldots,y_N\bigr)\,,
\end{multline}
provided the initial datum $\phi_0$ (at all times set to zero) has the corresponding symmetry: For all integers $M,N\geq 0$ and all permutations $\pi$ of $\{1,\ldots,M\}$ and $\rho$ of $\{1,\ldots,N\}$,
\begin{multline}\label{0permutation}
\phi_{0,r_{\pi(1)}\ldots r_{\pi(M)},s_{\rho(1)}\ldots s_{\rho(N)}}
\bigl(\vx_{\pi(1)},\ldots,\vx_{\pi(M)},\vy_{\rho(1)},\ldots,\vy_{\rho(N)}\bigr) =\\ 
(-1)^\pi \phi_{0,r_1\ldots r_M,s_1\ldots s_N}\bigl(\vx_1,\ldots,\vx_M,\vy_1,\ldots,\vy_N\bigr)\,.
\end{multline}
\end{ass}

The derivation is given in Section~\ref{sec:proof_permutation}; alternatively, it follows from Assertion~\ref{ass:Heisenberg_Picture} in Section~\ref{sec:operatorvaluedfields} and the canonical (anti-)commutation relations. Note the difference between $x_j$ (space-time point) and $\vx_j$ (space point); that is, time variables are also permuted in \eqref{permutation} but not in \eqref{0permutation}. The symmetry \eqref{0permutation} is, of course, the usual permutation symmetry for the case that $x$-particles are fermions and $y$-particles are bosons. Note also that the symmetry \eqref{permutation} of $\phi$ directly implies that the 1-time wave function $\psi$ obtained from $\phi$ by setting all time variables equal as in \eqref{psitphi} satisfies the usual symmetry \eqref{0permutation}, a fact that could also be concluded from the observation that $\psi$ obeys the 1-time Schr\"odinger equation \eqref{Schr1Hdef} and has the same initial datum $\psi_0=\phi_0$ as $\phi$. In much the same way, \eqref{permutation} also implies that $\psi_\Sigma$ as in \eqref{psiSigmadef} has the usual permutation symmetry, i.e., the analog of \eqref{0permutation} on $\Sigma$; put differently, $\psi_\Sigma$ lies in the product of Fock spaces over $\Sigma$, i.e., in the space $\Fock_{\Anti} (\Hilbert_{1,\Sigma}) \otimes \Fock_{\Sym}(\Hilbert_{1,\Sigma})$ with $\Fock_{\Anti},\Fock_{\Sym}$ the fermionic and bosonic Fock spaces and $\Hilbert_{1,\Sigma}$ the 1-particle Hilbert space for $\Sigma$ (as described around \eqref{scpHilbertSigma}--\eqref{scpHilbert1Sigma} below).

Marx~\cite{Marx:1972} already expected that \eqref{permutation} is the correct expression of fermionic and bosonic symmetry. He further observed that, for any fixed $M$ and $N$ and any fixed choice of times $x_1^0,\ldots,x_M^0,y_1^0,\ldots,y_N^0$, $\phi(x_1,\ldots,x_M,y_1,\ldots,y_N)$ as a function of the spatial coordinates will in general be asymmetric (i.e., neither symmetric nor anti-symmetric). He suggested to introduce modified Fock spaces that do not require any particular symmetry type but contain asymmetric functions. However, this suggestion is not compelling. It is inspired by regarding $\phi_T$ as the analog of $\psi_t$, a thought we have discussed critically in Remark~\ref{rem:psit} in Section~\ref{sec:MT-em-ab} above. In contrast, since the possible $\psi_\Sigma$ do obey the permutation symmetry analogous to \eqref{0permutation}, their Fock space is the ordinary Fock space.

\subsection{Behavior Under Lorentz Transformations}
\label{sec:LorentzTrafo}

We now describe how the system of multi-time equations \eqref{multi12} behaves under Lorentz transformations. The upshot will be that (roughly speaking) the multi-time equations are invariant if we transform $g$ and $G$ appropriately; put differently, the equations would be invariant if $g$ and $G$ were, but (unlike the 4-vector of Dirac gamma matrices $\gamma^\mu$) they are not. The reason why we chose such $g$ and $G$ is because then the creation and annihilation terms can be written down in an easy way. Fully Lorentz-invariant multi-time equations can presumably be written down if one takes the $y$ particles to have spin~$1$. We refrained from doing so only because of the aforementioned problems with the position representation for photons.

To study the transformation behavior in detail, it is useful to consider the generalization of \eqref{multi12} provided by Remark~\ref{rem:g3}. Furthermore, just like the 1-particle Dirac equation can be written either in the Hamiltonian form $i\partial_0 \psi = (-i\valpha\cdot\nabla +\beta m)\psi$ or in the manifestly covariant form $i\gamma^\mu \partial_\mu \psi = m\psi$, it is useful to re-write the multi-time equations in the same way, using the notation $\partial_{j\mu}=\partial/\partial x_j^\mu$, $\partial_{k\mu}=\partial/\partial y_k^\mu$ (so that the symbols $j$ and $k$ convey whether the variable is $x$ or $y$), $\gamma_j^\mu$ being the $\gamma^\mu$ matrix acting on $r_j$ and $\gamma_k^\mu$ the one acting on $s_k$, and with implicit summation over repeated space-time indices (such as $\mu$) but not over particle labels (such as $j$ or $k$):
\begin{subequations}\label{multi34}
\begin{align}\label{multi3}
i \gamma_j^\mu\partial_{j\mu}\phi(x^{4M}, y^{4N}) &= m_x \phi(x^{4M}, y^{4N}) \nonumber\\
&\quad + \sqrt{N+1}  \!\! \sum_{s_{N+1},r'_j=1}^4 \!\!\! (\tilde{g}^+)_{~~~~~~~r_j}^{s_{N+1}r'_j} \, \phi_{r'_js_{N+1}}\bigl(x^{4M}, (y^{4N}, x_j)\bigr)
\nonumber \\
& \quad + \frac{1}{\sqrt{N}} \sum_{k=1}^N \sum_{r'_j=1}^4 \tilde{G}_{s_kr_j}^{~~~\:r'_j}(y_k - x_j) \, \phi_{r'_j\widehat{s_k}}(x^{4M}, y^{4N} \backslash y_k)  \\[3mm] 
\label{multi4}
i \gamma_k^\mu\partial_{k\mu}\phi(x^{4M}, y^{4N}) &= m_y\phi(x^{4M}, y^{4N})
\end{align}
\end{subequations}
with
\begin{subequations}
\begin{align}
\tilde{G}_{sr}^{~~r'}(y)&=\sum_{r''=1}^4 (\gamma^0)_{rr''} G_{sr''r'}(y)\label{tildeGdef}\\
(\tilde{g}^+)_{~~\:r}^{sr'}&=\sum_{r''=1}^4 (\gamma^0)_{rr''} \, g^*_{sr'r''}\,.\label{tildeg+def}
\end{align}
\end{subequations}
In the notation $\tilde{G}$ and $\tilde{g}^+$, upper spin indices refer to $S^*$, the dual space of $S$, while lower ones refer to $S$. The notation $^+$ in $\tilde{g}^+$ will be elucidated later in this subsection. 
In this form, obtained from the multi-time equations \eqref{multi12} with the changes \eqref{g3changes1} and \eqref{g3changes2} of Remark~\ref{rem:g3} by moving the free Hamiltonian to the left-hand side and multiplying by $\gamma_j^0$ respectively $\gamma_k^0$, the differential operators are manifestly covariant.

Let us now consider a Lorentz transformation $\Lambda$ from the Lorentz frame $L$ to the Lorentz frame $\underline{L}$; if a 4-vector has components $u^\mu$ relative to $L$ then we write the components relative to $\underline{L}$ as
\be
\underline{u}^{\underline{\mu}} = \Lambda^{\underline{\mu}}_{~\:\mu} u^\mu\,.
\ee
We write $\sM$ for Minkowski space-time and $\tilde\Lambda_{\underline{s}}^{~\:s}$ for the corresponding basis change in spin space $S$ (as every Lorentz frame in $\sM$ is naturally associated with a basis in $S$); that is, $\tilde\Lambda$ is the action of $\Lambda$ on $S$. We can think of $\phi$ as a function on the spacelike configurations in $\Gamma(\sM)^2$ with values in the appropriate spin space, $\phi(x^{4M},y^{4N})\in S^{\otimes (M+N)}$. The nature of $\phi$ then determines its transformation behavior; namely, a Lorentz transformation amounts to giving different coordinates to points in $\sM$ and choosing a different basis in spin space $S$. Put differently, the transform $\underline{\phi}$ reads (we write $\phi$ again for the original coordinate representation)
\be
\underline{\phi}_{\underline{r}_1\ldots \underline{s}_N}(\underline{x}_1,\ldots,\underline{y}_N) = 
\sum_{r_1,\ldots,s_N=1}^4 \tilde\Lambda_{\underline{r}_1}^{~~r_1}\cdots \tilde\Lambda_{\underline{s}_N}^{~~s_N} \, 
\phi_{r_1\ldots s_N}\bigl( \Lambda^{-1}(\underline{x}_1),\ldots, \Lambda^{-1}(\underline{y}_N) \bigr)\,.
\ee
It is clear from the form of \eqref{multi34} that $\underline{\phi}$ satisfies \eqref{multi34} with the appropriate transforms of $\tilde{G}$, and $\tilde{g}^+$, i.e.,
\begin{subequations}
\begin{align}
\underline{\tilde{G}}_{\underline{s}\underline{r}}^{~~~\underline{r}'}(\underline{x})&=\sum_{r,r',s=1}^4 \tilde\Lambda_{\underline{s}}^{~\:s}\;\tilde\Lambda_{\underline{r}}^{~\:r}\;\tilde\Lambda^{\underline{r}'}_{~r'}\:\:\tilde{G}_{sr}^{~~r'}\!\bigl(\Lambda^{-1}(\underline{x})\bigr)\\
(\underline{\tilde{g}^+})^{\underline{s} \underline{r}'}_{~~~\underline{r}}&= \sum_{r,r',s=1}^4 \tilde\Lambda^{\underline{s}}_{~\:s}\;\tilde\Lambda^{\underline{r}'}_{~\:r'}\;\Lambda_{\underline{r}}^{~\:r}\:(\tilde{g}^+)^{sr'}_{~~r}\,.
\end{align}
\end{subequations}
That is, \eqref{multi34} is Lorentz-invariant if we regard $\tilde{G}$ as a function $\sM\to S\otimes S \otimes S^*$ and $\tilde{g}^+$ as an element of $S^*\otimes S^* \otimes S$. Of course, these objects are themselves not Lorentz-invariant (unlike, for example, the 4-vector $\gamma^\mu$ of Dirac gamma matrices); in particular, they may prefer one Lorentz frame over others.

Furthermore, $\tilde{G}$ and $\tilde{g}^+$ can be obtained from an element $\tilde{g}\in S\otimes S\otimes S^*$, as we explain now. 

As a preparation, we need a basic fact about the Green functions: Suppose we pick a spinor $g\in S$ (with components $g_s$ relative to the spin basis corresponding to the Lorentz frame $L$) and form the Green function starting from $g_s$, i.e., $G_s(x)$ satisfies
\be\label{Green1}
i\gamma^\mu\partial_\mu G = mG
\text{ and }
G_s(0,\vx)=g_s\,\delta^3(\vx)\,.
\ee
When we Lorentz transform $G$ and obtain $\underline{G}$ then $\underline{G}$ is again a Green function,
\be\label{Green2}
i\gamma^\mu\partial_\mu \underline{G}=m\underline{G}
\text{ and }
\underline{G}_s(0,\vx)=g'_s\,\delta^3(\vx)\,,
\ee
but $g'$ is not the transform $\underline{g}$ of $g$ (in the sense $\underline{g}=\tilde\Lambda g$ in which Lorentz transformations act on spin space). Rather, $g'$ is the transform of $\Lambda^{\underline{0}}_{~\mu}\gamma^\mu \gamma^0\, g$ (where $\Lambda^{\underline{0}}_{~\mu}$ can be regarded as the new timelike basis vector expressed in the old coordinates). Put differently, if we write $g$ as $\gamma^0 \tilde{g}$, then the components of $g'$ in the new spin basis can be written as $\gamma^0 \tilde\Lambda \tilde{g}$. This follows from the known fact (see, e.g., Proposition~6.1 in \cite{Fin13}; it also follows from Theorem~1.2 in \cite{thaller:1992}) that $G(x)= k(x)\, \gamma^0\, g = k(x)\, \tilde{g}$, where $k(x)$ is a certain Lorentz-invariant $S\otimes S^*$-valued distribution. (Note that $\gamma^0\gamma^0=1$.) Thus, the initial spinor of $G$ at the origin can be characterized in a Lorentz-covariant way by specifying $\tilde{g}$ (rather than $g$).

This carries over to our $S\otimes S\otimes S^*$-valued Green functions as follows. Let
\be\label{tildegdef}
\tilde{g}_{sr}^{~~r'}=\sum_{s',r''=1}^4(\gamma^0)_{ss'}(\gamma^0)_{rr''}g_{s'r''r'}\,.
\ee
This defines an element of $S\otimes S \otimes S^*$. Then $\tilde{G}_{sr}^{~~r'}\!(y)$ is the Green function (in the variable $y$ and the spin index $s$) with initial spinor covariantly characterized by $\tilde{g}$. Also $\tilde{g}^+$ can be obtained in a covariant way from $\tilde{g}$, as follows. While the inner product in spin space, $\psi^\dagger \chi$, is not Lorentz invariant, a certain indefinite bilinear form is; it is usually denoted by $\overline\psi \chi$, and is given by $\psi^\dagger \gamma^0 \chi$. This form defines, in a Lorentz-invariant way, a bijective mapping $S\to S^*$, which can be expressed as $\psi \mapsto \psi^\dagger \gamma^0$ and which we denote by ${}^+$. This mapping is conjugate-linear. Whenever, for $i=1,2$, $S_i$ and $T_i$ are complex vector spaces and $\varphi_i:S_i\to T_i$ is a conjugate-linear mapping, then $\varphi_1\otimes \varphi_2$ denotes the unique conjugate-linear mapping $S_1\otimes S_2\to T_1\otimes T_2$ that will map, for any $s_1\in S_1$ and $s_2\in S_2$, $s_1\otimes s_2$ to $\varphi_1(s_1)\otimes \varphi_2(s_2)$; correspondingly, ${}^+\otimes {}^+\otimes ({}^+)^{-1}$ is a conjugate-linear, Lorentz-invariant bijection $S\otimes S\otimes S^*\to S^*\otimes S^* \otimes S$, which for brevity we denote again by ${}^+$; this is the relation between $\tilde{g}$ and $\tilde{g}^+$. In coordinates, 
\begin{subequations}
\begin{align}
(\psi^+)^s&=\sum_{s'}(\gamma^0)_{s's}\psi^*_{s'}\,,\\
\psi_s&=\sum_{s'}(\gamma^0)_{ss'}(\psi^+)^{s'*}
\end{align}
(because $\gamma^0$ is self-adjoint and self-inverse), and thus
\be
(\tilde{g}^+)^{sr'}_{~~\:r}=\sum_{r'',r''',s'}(\gamma^0)_{s's}(\gamma^0)_{r'''r'}(\gamma^0)_{rr''}\tilde{g}_{s'r'''}^{~~~~\:r''*}\,,
\ee
\end{subequations}
in agreement with \eqref{tildeg+def} and \eqref{tildegdef}.

To sum up, if we think of the spin space $S$ as (not $\CCC^4$ but) defined in a covariant way from Minkowski space, then, once an element $\tilde{g}$ of $S\otimes S\otimes S^*$ has been chosen, $\tilde{g}^+$ is obtained from $\tilde{g}$ through the Lorentz-invariant operation ${}^+$; $\tilde{G}$ is obtained as the Green functions with initial spinor covariantly characterized by $\tilde{g}$; and \eqref{multi34} is an invariant system of equations.

\subsection{Generalization to Curved Space-Time}
\label{sec:curved}

The multi-time system \eqref{multi12}, in the generalized version provided by Remark~\ref{rem:g3} and summarized by \eqref{multi34}, can be generalized straightforwardly to curved space-time $(\sM,g)$.\footnote{We need to make the technical assumption that $(\sM,g)$ permits a \emph{spin structure} and, in case the spin structure is not unique, that one has been chosen; see, e.g., \cite[Section~1.5]{PR84} for discussion. For existence and uniqueness of solutions to the multi-time system, further assumptions (such as global hyperbolicity) may be necessary.} In this setting, $\phi$ is defined on the spacelike configurations in $\Gamma(\sM)^2$ with values
\be
\phi(x_1,\ldots,x_M,y_1,\ldots,y_N)\in
 S_{x_1}\otimes \cdots \otimes S_{x_M}\otimes S_{y_1}\otimes \cdots \otimes S_{y_N}\,,
\ee
where $S_x$ is a fiber of the bundle $S$ of spin spaces, a vector bundle over the base manifold $\sM$. This can be expressed using the notation $A\boxtimes B$ for the vector bundle over the base manifold $\sA\times \sB$ (obtained from vector bundles $A$ over $\sA$ and $B$ over $\sB$) whose fiber at $(a,b)\in\sA\times\sB$ is
\be
(A\boxtimes B)_{(a,b)} = A_a\otimes B_b\,,
\ee
and correspondingly $A^{\boxtimes n}$ for $A\boxtimes A \boxtimes \cdots \boxtimes A$ with $n$ factors. Then the $(M,N)$-particle sector of $\phi$ is a cross-section of the vector bundle $S^{(M,N)}=S^{\boxtimes M}\boxtimes  S^{\boxtimes N}$ over $\sM^{M+N}$. The equations \eqref{multi34} need only minor changes and re-interpretation of symbols. The free Dirac operators have to be understood appropriately; namely, $\partial_{j\mu}$ is now the covariant derivative on $S$ corresponding to the connection naturally associated with the metric of $\sM$ \cite[Section~4.4]{PR84}, and correspondingly on $S^{\boxtimes (M+N)}$. Spin indices $r$ and $s$ refer to the appropriate spin space $S_x$. The gamma matrices are understood as a cross-section of $T\sM\otimes S\otimes S^*$ (where $T\sM$ denotes the tangent bundle and $*$ the dual space) that naturally comes with $S$, and $\gamma_j^\mu$ in \eqref{multi3} [respectively, $\gamma_k^\mu$ in \eqref{multi4}] is understood as $\gamma_j^\mu(x_j)$ [respectively, $\gamma_k^\mu(y_k)$]. The coefficient $\tilde{g}_{sr'}^{~~r}$ gets replaced in \eqref{multi3} by a cross-section $\tilde{g}_{sr'}^{~~r}(x_j)$ of the bundle $S\otimes S\otimes S^*$. Similarly, $\tilde{G}(y-x)$ in \eqref{multi3} gets replaced by $\tilde{G}(y,x)$, which is the appropriate Green function, namely the solution of the free Dirac equation in $y$ with initial spinor at $x$ covariantly characterized by $\tilde{g}(x)$.

Our consistency proof (of Assertion~\ref{ass:consistency}) still applies.

\section{Operator-Valued Fields and Multi-Time Wave Functions}
\label{sec:operatorvaluedfields}

\begin{ass}\label{ass:Heisenberg_Picture}
For the emission--absorption model, let $\phi$ be the multi-time wave function, i.e., a solution to \eqref{multi12}, let $H$ be the single-time Hamiltonian as in \eqref{Hdef}, let $a_r(t,\vx)=e^{iHt}a_r(\vx)e^{-iHt}$, likewise $b_s(t,\vy)$, and let $\psi_0$ be the initial wave function, i.e., equal to $\phi$ with all time variables set to zero. Then, for any $(x_1,\ldots, x_M,y_1,\ldots, y_N)\in\sS_{xy}$,
\begin{align}
&\phi_{r_1\ldots r_M,s_1\ldots s_N}(x_1,\ldots,x_M,y_1,\ldots,y_N) \nonumber\\
&\quad= \frac{(-1)^{M(M-1)/2}}{\sqrt{M!N!}}\Bscp{\emptyset}{a_{r_1}(x_1)\cdots a_{r_M}(x_M)b_{s_1}(y_1)\cdots b_{s_N}(y_N)\Big|\psi_0}\,,\label{phiabPsi}\\
\intertext{and if the configuration is collision-free (i.e., if the $x_j$, $y_k$ are pairwise distinct), then}
&\phi_{r_1\ldots r_M,s_1\ldots s_N}(x_1,\ldots,x_M,y_1,\ldots,y_N) \nonumber\\
 &\quad= \frac{(-1)^{M(M-1)/2}}{\sqrt{M!N!}} \Bscp{\emptyset}{\Phi_{x,r_1}(x_1)\cdots \Phi_{x,r_M}(x_M)\Phi_{y,s_1}(y_1)\cdots \Phi_{y,s_N}(y_N)\Big|\psi_0}\,,\label{phiFPsi}
\end{align}
where $|{\emptyset}\rangle$ is the Fock vacuum state and
\be\label{Fabdef}
\Phi_{x,r}(x)=a_r(x)+a_r^\dagger(x)\,, \quad \Phi_{y,s}(y)=b_s(y)+b_s^\dagger(y)
\ee
are the field operators. 
\end{ass}

The derivation is given in Section \ref{proofs:Heisenberg_Picture}.

The combinatorial factor in \eqref{phiabPsi} can be understood as follows. Since we are using wave functions (such as $\phi$) that are functions of \emph{ordered} configurations (while the physical configurations are unordered), normalization of $\phi$ requires to shrink the expression that would be natural on the space of unordered configurations by a factor $(M!N!)^{-1/2}$. If we used only configuration spaces of unordered configurations, these combinatorial factors would be absent; see \cite[section 2.4]{GTTZ:2013} for further discussion of the spaces of ordered and unordered configurations.

\section{Tomonaga--Schwinger Approach and Multi-Time Wave Functions}
\label{sec:TS}

We begin with the fact that the multi-time equations naturally provide a unitary evolution for the wave functions $\psi_\Sigma$ as in \eqref{psiSigmadef}.

\begin{ass}\label{ass:unitary_HS_evolution}
In the emission--absorption model,
for any spacelike hypersurfaces $\Sigma,\Sigma'$, the multi-time equations \eqref{multi12} define a unitary evolution $U_{\Sigma\to\Sigma'}: \Hilbert_{\Sigma}\to \Hilbert_{\Sigma'}$. That is, initial data $\psi$ on $\Sigma$ select a unique solution $\phi$ of \eqref{multi12}, whose restriction to $\Sigma'$ is $U_{\Sigma\to\Sigma'}\psi$ for a unitary isomorphism $U_{\Sigma\to\Sigma'}$.
\end{ass}

The derivation is given in Section~\ref{proofs:Tomonaga-Schwinger}. Here, the Hilbert space $\Hilbert_\Sigma$ consists of functions $\psi$ on $\Gamma(\Sigma)\times \Gamma(\Sigma)$ with values $\psi(x^{4M},y^{4N})\in(\CCC^4)^{\otimes M}\otimes (\CCC^4)^{\otimes N}$, anti-symmetric in the $x$ (and their spin indices) and symmetric in the $y$ (and their spin indices); the inner product is given by
\begin{multline}\label{scpHilbertSigma}
\scp{\psi}{\chi}_{\Hilbert_\Sigma} = \sum_{M,N=0}^\infty \: \int\limits_{\Sigma^{M+N}} d^3x_1\cdots d^3x_M d^3y_1\cdots d^3y_N \:\times\\
\times \: \psi^*(x^{4M},y^{4N}) \Bigl[ \gamma^0 \gamma^{\mu_1} n_{\mu_1}(x_1) \otimes \cdots \otimes \gamma^0 \gamma^{\mu_{M+N}} n_{\mu_{M+N}}(y_N)\Bigr] \chi(x^{4M},y^{4N})
\end{multline}
with $\psi^*$ the conjugate-transpose of $\psi$, $d^3x$ referring to the invariant 3-volume measure defined by the Riemannian 3-metric on $\Sigma$, and $n_\mu(x)$ again the future-pointing unit normal vector to $\Sigma$ at $x$.
Put differently, 
\be\label{HilbertSigmadef}
\Hilbert_\Sigma = \Fock_{\Anti} (\Hilbert_{1,\Sigma}) \otimes \Fock_{\Sym}(\Hilbert_{1,\Sigma})
\ee
with $\Fock$ the Fock space and $\Hilbert_{1,\Sigma}$ the 1-particle Hilbert space for $\Sigma$ consisting of functions $u:\Sigma\to\CCC^4$ with inner product
\be\label{scpHilbert1Sigma}
\scp{u}{v}_{\Hilbert_{1,\Sigma}} = \int_\Sigma d^3x \, u^*(x) \, \gamma^0 \, \gamma^\mu \, n_\mu(x) \, v(x)\,.
\ee
Note that $\gamma^0\, \gamma^\mu\, n_\mu(x)$ is a positive-definite self-adjoint matrix (while $\gamma^\mu\, n_\mu(x)$ is typically not self-adjoint). We now elucidate the Tomonaga--Schwinger approach.

\subsection{The Tomonaga--Schwinger Approach}

Taking a more general perspective and considering not only our emission--absorption model but rather arbitrary QFTs, we take for granted that with every spacelike hypersurface $\Sigma$ there is associated a Hilbert space $\Hilbert_\Sigma$, and with any two $\Sigma,\Sigma'$ a unitary isomorphism $U_{\Sigma\to\Sigma'}:\Hilbert_\Sigma\to\Hilbert_{\Sigma'}$ such that $U_{\Sigma'\to\Sigma''}U_{\Sigma\to\Sigma'}=U_{\Sigma\to\Sigma''}$, and $U_{\Sigma\to\Sigma}$ is the identity. To express the time evolution by means of unitary operators within a \emph{fixed} Hilbert space $\tilde\Hilbert$, as desired in the Tomonaga--Schwinger approach, one needs to identify each $\Hilbert_\Sigma$ with $\tilde\Hilbert$; but this amounts to an identification between $\Hilbert_\Sigma$ and $\Hilbert_{\Sigma'}$ for any $\Sigma,\Sigma'$. There are only two natural ways at hand to obtain such an identification: the time evolution $U_{\Sigma\to\Sigma'}$ itself, and the \emph{free} (non-interacting) time evolution $F_{\Sigma\to\Sigma'}$ (which is also a unitary isormorphism $\Hilbert_\Sigma\to\Hilbert_{\Sigma'}$). If we use $U_{\Sigma\to\Sigma'}$ then the time evolution in $\tilde\Hilbert$ is trivial (Heisenberg picture); if we use $F_{\Sigma\to\Sigma'}$ then the time evolution in $\tilde\Hilbert$ represents the \emph{interaction picture}.\footnote{One could think of the Schr\"odinger picture in non-relativistic quantum mechanics as exemplifying a third way as follows. There, the only hypersurfaces we consider are $\Sigma=\{t=\mathrm{const.}\}$ in a fixed Lorentz frame $L$, and the use of a fixed Hilbert space $\tilde\Hilbert$ can be regarded as corresponding to identifications $I_{\Sigma\to\Sigma'}:\Hilbert_\Sigma\to\Hilbert_{\Sigma'}$ that, using that $\Hilbert_\Sigma$ consists of functions on $\Sigma^N$ or $\Gamma(\Sigma)$, arise from isometries $J:\Sigma\to\Sigma'$, where $J(x)$ is simply the point on $\Sigma'$ with the same 3 space coordinates (in the frame $L$) as $x$. However, a similar strategy is not available for two arbitrary (curved) spacelike hypersurfaces because they are not isometric. (One might have hoped that it helps that they are diffeomorphic, with one particular diffeomorphism given by mapping $x\in\Sigma$ to the point on $\Sigma'$ with the same 3 space coordinates in $L$. However, the corresponding mapping $\Hilbert_\Sigma\to\Hilbert_{\Sigma'}$ then is not unitary.)} That is, once we choose some spacelike hypersurface $\Sigma_0$ and some unitary mapping $\tilde{U}_{\Sigma_0}$ for identifying $\Hilbert_{\Sigma_0}$ with $\tilde\Hilbert$ then every other $\Hilbert_\Sigma$ gets identified with $\tilde\Hilbert$ via $\tilde{U}_\Sigma= \tilde{U}_{\Sigma_0} F_{\Sigma\to\Sigma_0}$, the wave function $\psi_\Sigma\in\Hilbert_\Sigma$ gets represented by the vector $\tilde\psi_\Sigma=\tilde{U}_\Sigma \psi_\Sigma$ in $\tilde\Hilbert$, and the time evolution $\tilde{U}_{\Sigma\to\Sigma'}:\tilde\Hilbert\to\tilde\Hilbert$ that maps $\tilde\psi_{\Sigma}$ to $\tilde\psi_{\Sigma'}$ is given by
\be\label{tildeUSigmaSigma'}
\tilde{U}_{\Sigma\to\Sigma'} 
= \tilde{U}_{\Sigma'} U_{\Sigma\to\Sigma'} \tilde{U}_{\Sigma}^{-1}
= \tilde{U}_{\Sigma_0} F_{\Sigma'\to\Sigma_0} U_{\Sigma\to\Sigma'} F_{\Sigma_0\to\Sigma} \tilde{U}_{\Sigma_0}^{-1}\,.
\ee
It is the time evolution $\tilde{U}_{\Sigma\to\Sigma'}$ that the Tomonaga--Schwinger equation \eqref{TS} is intended to characterize, for infinitesimally neighboring $\Sigma,\Sigma'$. For the evolution to be Lorentz invariant, $\Hd_{I}(x)$ must be a Lorentz scalar.

A solution of the Tomonaga--Schwinger equation is a function $\tilde\psi$ on the ($\infty$-dimensional) set of all spacelike hypersurfaces with values in $\tilde\Hilbert$. Such a solution exists for arbitrary initial datum $\tilde\psi_{\Sigma_0}$ if and only if the consistency condition
\be\label{TSconsistency}
\bigl[ \Hd_I(x),\Hd_I(y) \bigr]=0 
\text{ whenever }x\sim y
\ee
holds ($\sim$ means spacelike separated). If the consistency condition is violated then the Tomonaga--Schwinger equation \eqref{TS} can still be solved for any foliation of space-time, but on two different foliations interpolating between $\Sigma_0$ and $\Sigma$, \eqref{TS} may lead to different versions of $\tilde\psi_\Sigma$, starting from the same initial datum $\tilde\psi_{\Sigma_0}$. 

As mentioned in Remark~\ref{rem:consistency} in Section~\ref{sec:MT-em-ab}, also for multi-time equations of the form
\be\label{phiHjabs}
i\frac{\partial\phi}{\partial t_j}=H_j\phi\,,
\ee
there is a consistency condition. The simplest case, discussed extensively in \cite{pt:2013a}, is to consider a fixed number $N$ of time variables and solutions of the form $\phi:\RRR^N\to\Hilbert$ with $\RRR^N$ the space spanned by the $N$ time axes and $\Hilbert$ some Hilbert space (containing, say, functions of $\vx_1,\ldots,\vx_N$). Then $H_j=H_j(t_1,\ldots,t_N)$ are operators on $\Hilbert$, and the \emph{consistency condition} reads
\be
\bigl[H_j,H_k\bigr] -i\frac{\partial H_k}{\partial t_j}+i\frac{\partial H_j}{\partial t_k}=0
\ee
or, more compactly,
\be\label{MTconsistency}
\biggl[i\frac{\partial}{\partial t_j}-H_j,i\frac{\partial}{\partial t_k}-H_k\biggr]=0\,.
\ee
If it is satisfied, then there is a unique joint solution of \eqref{phiHjabs} for every initial datum $\phi(0,\ldots,0)\in\Hilbert$. If it is not satisfied, then it is still possible to define the evolution along a path in the space $\RRR^N$ spanned by the time axes, but two different paths from $(0,\ldots,0)$ to $(t_1,\ldots,t_N)$ may lead to different results for $\phi(t_1,\ldots,t_N)$; this situation is analogous to the foliation-dependence of $\tilde\psi_\Sigma$ described above in case \eqref{TSconsistency} is violated.

In Section~\ref{sec:TSfromMT} we formulate that the Tomonaga--Schwinger equation holds in the emission--absorption model, and in Section~\ref{sec:MTfromTS} we describe under which conditions the Tomonaga--Schwinger equation implies the existence of a multi-time wave function $\phi$.

\subsection{Tomonaga--Schwinger Equation from Multi-Time Wave Function}
\label{sec:TSfromMT}

Let us come back to the emission--absorption model. The free time evolution defines a unitary isomorphism $F_{\Sigma\to\Sigma'}:\Hilbert_\Sigma\to\Hilbert_{\Sigma'}$ for any two spacelike hypersurfaces $\Sigma,\Sigma'$. 
Indeed, it is known \cite{dimock:1982} that for a single particle, the free Dirac equation defines a unitary isomorphism $F^{(1)}_{\Sigma\to\Sigma'}:\Hilbert_{1,\Sigma}\to\Hilbert_{1,\Sigma'}$ for any two spacelike hypersurfaces $\Sigma,\Sigma'$. Explicitly, $F^{(1)}_{\Sigma\to\Sigma'}\psi_{1,\Sigma}$ can be obtained by solving the Dirac equation on $\RRR^4$ with initial data $\psi_{1,\Sigma}$ on $\Sigma$ and then restricting the solution to $\Sigma'$. From $F^{(1)}_{\Sigma\to\Sigma'}$, we obtain
\be
F_{\Sigma\to\Sigma'}= \Bigl(\bigoplus_{M=0}^\infty F^{(1,x)\otimes M}_{\Sigma\to\Sigma'}\Bigr) 
\otimes \Bigl( \bigoplus_{N=0}^\infty F^{(1,y)\otimes N}_{\Sigma\to\Sigma'}\Bigr)\,,
\ee
where $F^{(1,x)}$ and $F^{(1,y)}$ involve different masses $m_x,m_y$.

Let $\Sigma_0$ be the hypersurface $\{x^0=0\}$. 
Let $N_x(\vx)$ be the particle number density operator for the $x$-particles acting on $\Hilbert_{\Sigma_0}$,
\be\label{Ndef}
N_x(\vx) = \sum_{r=1}^4 a^\dagger_r(\vx) \, a_r(\vx)
\ee
with $a_r(\vx)$ as in \eqref{adef}; $N_x(\vx)$ is equivalently characterized by
\be\label{Ndeltapsi}
N_x(\vx) \psi(x^{3N},y^{3m}) = \sum_{j=1}^N \delta^3(\vx_j-\vx) \, \psi(x^{3N},y^{3m})\,.
\ee
We are now ready to formulate 

\begin{ass}\label{ass:Tomonaga-Schwinger}
Let $\phi$ evolve according to the multi-time equations \eqref{multi12} of the emission--absorption model, set
\be
\psi_\Sigma(x_1,\ldots,x_M,y_1,\ldots,y_N)=\phi(x_1,\ldots,x_M,y_1,\ldots,y_N)
\ee
for any $x_1,\ldots,x_M,y_1,\ldots,y_N\in\Sigma$, and let $\tilde\psi_\Sigma$ be the interaction picture of $\psi_\Sigma$,
\be\label{tildepsidef}
\tilde\psi_\Sigma = F_{\Sigma\to\Sigma_0}\psi_\Sigma\,.
\ee
Then $\tilde\psi$ satisfies the Tomonaga--Schwinger equation \eqref{TS} with $\Hd_I(x)$ the interaction Hamiltonian density in the interaction picture, which is\footnote{Of course, this expression is not covariant, as our emission--absorption model is not fully covariant.}
\be\label{HdIdef2}
\Hd_I(t,\vx)
= e^{iH^\free t} \Bigl( N_x(\vx) \otimes \sum_{s=1}^4 \bigl( g_s^*\, b_s(\vx)+g_s \, b^\dagger_s(\vx) \bigr) \Bigr) e^{-iH^\free t}\,.
\ee
\end{ass}

The derivation is given in Section~\ref{proofs:Tomonaga-Schwinger}.

\subsection{Multi-Time Wave Function from Tomonaga--Schwinger Equation}
\label{sec:MTfromTS}

Recall that a family $(\psi_\Sigma)_\Sigma$ of wave functions on every spacelike hypersurface $\Sigma$ corresponds to a multi-time wave function $\phi$ iff, for every configuration $q$ on $\Sigma$ and every other $\Sigma'$ also containing $q$,
\be\label{psiSigmaprimeSigma}
\psi_{\Sigma'}(q)=\psi_\Sigma(q)\,.
\ee
The next assertion shows that this condition follows from the Tomonaga--Schwinger equation under assumptions that we think are usually satisfied. 

To begin with, $\tilde\psi_\Sigma$ as given by the Tomonaga--Schwinger equation is a vector in an abstract Hilbert space $\tilde{\Hilbert}$, and in order to even be able to talk about $\psi_\Sigma(q)$, i.e., to be able to evaluate a wave function at a particular configuration $q$, we need further information about how $\tilde{\Hilbert}$ connects to particle configurations. To this end, we assume first that $\Hilbert_\Sigma$ is a subspace of a continuous tensor product,
\be\label{HilbertSigmaHilbertx}
\Hilbert_\Sigma\subseteq \widehat\Hilbert_\Sigma= \int^{\otimes}_{\Sigma} d^3x\, \Hilbert_x\,,
\ee
where $\int^{\otimes}$ means the continuous tensor product.\footnote{The symbol $\Hilbert_x$, which we used before for the Hilbert space of the $x$-particles (as distinct from the $y$-particles), now means the Hilbert space associated with the space-time point called $x$.} (We do not attempt here to give a definition of the continuous tensor product.) This is the case when $\Hilbert_\Sigma$ is a Fock space, or a tensor product of Fock spaces, each of some 1-particle Hilbert space $\Hilbert_{1,\Sigma}$ that is a space of $L^2$ functions on $\Sigma$ (with values in, e.g., some vector bundle $S$ of spin spaces $S_x$). Indeed, then
\be\label{Hilbert1SigmaSx}
\Hilbert_{1,\Sigma}=\int_\Sigma^{\oplus} d^3x \, S_x
\ee
and, since the (fermionic or bosonic) second quantization functor turns sums into products,
\be
\Fock(\Hilbert\oplus \Hilbert')=\Fock(\Hilbert)\otimes\Fock(\Hilbert')\,,
\ee
we have that
\be
\Fock\biggl(\int^\oplus d^3x\, S_x\biggr) = \int^\otimes d^3x \, \Fock(S_x)\,,
\ee
which is \eqref{HilbertSigmaHilbertx} with $\Hilbert_x = \Fock(S_x)$.

Furthermore, for a particle-position representation of elements of $\Hilbert_\Sigma$, we need that $\Hilbert_x$ consists of sectors corresponding to different particle numbers; that is, assuming we have $\ell$ different particle species, we demand that
\be\label{HilbertxHilbertxn}
\Hilbert_x = \bigoplus_{n_1\ldots n_\ell=0}^\infty \Hilbert_x^{(n_1\ldots n_\ell)}\,,
\ee
where some of the spaces $\Hilbert_x^{(n_1\ldots n_\ell)}$ may have dimension 0, so that not all numbers of particles are allowed. We demand further that the 0-particle sector is 1-dimensional, and that one vector in this sector is selected as the vacuum vector,
\be\label{dimHilbertx01}
\Hilbert_x^{(0\ldots 0)} = \CCC |\emptyset_x\rangle\,.
\ee
Again, this assumption holds for a tensor product of Fock spaces. We note that, in this setting, the vacuum vector $|\emptyset\rangle$ on $\Sigma$ is naturally defined as $|\emptyset\rangle = \int^\otimes_\Sigma d^3x\, |\emptyset_x\rangle$; that the number operator $N_i(x)$ on $\Hilbert_x$ for particle species $i$ has eigenvalue $n_i$ on the eigenspace $\oplus \Hilbert_x^{(n_1\ldots n_\ell)}$ with the sum taken over all values of $n_1,\ldots,n_{i-1},n_{i+1},\ldots,n_\ell$; and that, for any region $B$ of a spacelike hypersurface $\Sigma$, the number operator $N_i(B)$ on $\widehat\Hilbert_\Sigma$ for particle species $i$ can be defined by 
\be
N_i(B) = \int_B d^3x\, \Bigl(N_i(x) \otimes \int^\otimes_{\Sigma\setminus x} d^3x'\, I_{\Hilbert_{x'}}\Bigr)\,.
\ee
The $N_i(B)$ commute with each other because the integrands commute for different  $x$ (because they act non-trivially on different factors).

Then, for every $\psi_\Sigma\in\Hilbert_\Sigma$ and every configuration $q$ on $\Sigma$, we can define
\be\label{psiSigmaqdef}
\psi_\Sigma(q) := P_q \Bscp{ \int^{\otimes}_{\Sigma\setminus q} d^3x \, \emptyset_x}{\psi_\Sigma}
\ee
with $\scp{-}{-}$ a partial inner product (leaving out the factors $\Hilbert_x$ for $x\in q$), $P_{q}$ the projection (from $\otimes_{x\in q} \Hilbert_x$) to
\be\label{Hilbertqdef}
\Hilbert_{q}:= \bigotimes_{x\in q} \Hilbert_x^{(n_1(x,q)\ldots n_\ell(x,q))}\,,
\ee
and $n_i(x,q)$ the number of $i$-particles in $q$ at $x$. The multi-time wave function $\phi$ that we are about to construct will then take values $\phi(q)\in\Hilbert_q$.

\begin{ass}\label{ass:TStophi}
Consider a Tomonaga--Schwinger equation, i.e., let $\tilde{\Hilbert}$ be a Hilbert space and $\Hd_I(x)$ Hermitian operators on $\tilde\Hilbert$ with $[\Hd_I(x),\Hd_I(x')]=0$ whenever $x, x'$ are spacelike separated. Let $\Hilbert_\Sigma$ be given Hilbert spaces and $F_\Sigma:\tilde\Hilbert\to\Hilbert_\Sigma$ be unitary isomorphisms; define the ``free time evolution'' by $F_{\Sigma\to\Sigma'}= F_{\Sigma'}F_\Sigma^{-1}$. 
Assume \eqref{HilbertSigmaHilbertx}, \eqref{HilbertxHilbertxn}, and \eqref{dimHilbertx01}, as well as that
\be\label{HdIxHilbertx}
F_{\Sigma}\,\Hd_I(x)\, F_{\Sigma}^{-1}\text{ acts non-trivially only on }\Hilbert_x\,,
\ee
that
\be\label{HdIxHilbertx0}
F_{\Sigma}\,\Hd_I(x)\, F_{\Sigma}^{-1} |\emptyset_x\rangle = 0\,,
\ee
and that the free time evolution obeys \eqref{psiSigmaprimeSigma}, i.e., for any $q\subset \Sigma\cap\Sigma'$,
\be\label{freepsiSigmaprimeSigma}
F_{\Sigma\to\Sigma'}\psi_\Sigma(q)=\psi_\Sigma(q)\,.
\ee
Then the Tomonaga--Schwinger equation implies \eqref{psiSigmaprimeSigma} for the full time evolution. As a consequence, the $\psi_\Sigma$ fit together to form a multi-time wave function $\phi$.
\end{ass}

The derivation is given in Section~\ref{proofs:TStophi}.

As an illustration, let us explain why and how the assumptions are satisfied in our emission--absorption model. We took $\tilde{\Hilbert}=\Hilbert_{\Sigma_0}$; as mentioned in \eqref{HilbertSigmadef}, $\Hilbert_\Sigma$ is a tensor product of Fock spaces, with $\Hilbert_{1,\Sigma}$ (for both the $x$-particles and the $y$-particles) of the form \eqref{Hilbert1SigmaSx} with $S_x=\CCC^4$ a trivial vector bundle. As a consequence,
\be
\Hilbert_x=\Fock_{\Anti}(\CCC^4) \otimes \Fock_{\Sym}(\CCC^4)\,.
\ee
Note that $\dim \Fock_{\Anti}(\CCC^4) = 2^4$, while $\Fock_{\Sym}(\CCC^4)$ is $\infty$-dimensional. From the Fock space construction, we automatically obtain the decomposition of $\Hilbert_x$ into sectors corresponding to particle number, with a 0-particle sector of dimension 1 spanned by $|\emptyset_x\rangle$. Let us turn to the value space of $\phi$. Since collision configurations (i.e., those with more than one particle at a location) form a null set, we usually focus on configurations $q=(x_1,\ldots, x_M,y_1,\ldots, y_N)$ with $x_1,\ldots, y_N$ pairwise distinct; for such configurations, only the 0-particle and 1-particle sectors of $\Hilbert_x$ are relevant, $\Hilbert_x^{(0,0)}\cong \CCC,\Hilbert_x^{(1,0)}\cong \CCC^4$, and $\Hilbert_x^{(0,1)}\cong \CCC^4$; the value space of $\phi$ is
\be
\Hilbert_q = (\CCC^4)^{\otimes M}\otimes (\CCC^4)^{\otimes N}= \Hilbert_{x_1}^{(1,0)}\otimes \cdots \Hilbert_{x_M}^{(1,0)} \otimes \Hilbert_{y_1}^{(0,1)}\otimes \cdots \otimes \Hilbert_{y_N}^{(0,1)}\,,
\ee
in agreement with \eqref{Hilbertqdef}; for collision configurations, \eqref{Hilbertqdef} is still correct. As shown in \eqref{HdIxSigma} after Assertion~\ref{ass:axfree} below, 
\be\label{HdIxSigma2}
F_{\Sigma}\Hd_I(x)\,F_{\Sigma}^{-1}=  \sum_{r=1}^4 a^{\dagger}_{\Sigma,r}(x)a_{\Sigma,r}(x) \otimes \sum_{s=1}^4 \bigl( g_s^*\, b_{\Sigma,s}(x)+g_s \, b^\dagger_{\Sigma,s}(x) \bigr)\,,
\ee
with $a_\Sigma$ and $b_\Sigma$ the annihilation operators on $\Hilbert_\Sigma$. Note that $a^\dagger_{\Sigma,r}(x)$, creating a particle with wave function $g'\,\delta^3_\Sigma(\cdot-x)$ with the appropriate spinor $g'$, acts non-trivially only on $\Hilbert_x$; likewise with $a_{\Sigma,r}(x)$, $b^\dagger_{\Sigma,s}(x)$ and $b_{\Sigma,s}(x)$, and thus with $F_{\Sigma}\Hd_I(x)\,F_{\Sigma}^{-1}$. Note further that, since $a_{\Sigma,r}(x)|\emptyset_x\rangle =0$, the entire expression \eqref{HdIxSigma2} vanishes on $|\emptyset_x\rangle$. Furthermore, the free time evolution is such that the $\psi_\Sigma$ fit together to form a $\phi^\free$. Indeed, this follows from the corresponding fact for the free 1-particle evolution. Explicitly, the (many-particle multi-time wave function) $\phi^\free$ is defined, not only on the spacelike configurations $\sS$ but even on all configurations $\cup_{M,N=0}^\infty (\RRR^4)^{M+N}$, by
\be
\phi^\free(x_1,\ldots, x_M,y_1,\ldots, y_N) = \Bigl( e^{-iH^\free_{x_1}x_1^0} \cdots e^{-iH^\free_{y_N}y_n^0} \psi_0\Bigr)(\vx_1,\ldots,\vy_N)\,.
\ee
This shows that all assumptions of Assertion~\ref{ass:TStophi} are satisfied.

\section{Derivations of Assertions}
\label{sec:derivations}

The derivation of Assertion~\ref{ass:consistency} spans the subsections \ref{sec:non-spacelike}--\ref{sec:spacelike}.

\subsection{Inconsistency on Non-Spacelike Configurations}
\label{sec:non-spacelike}

We begin with the claim (contained in Assertion~\ref{ass:consistency}) that the multi-time system \eqref{multi12} is inconsistent on non-spacelike configurations if $m_y>0$ and $g^\dagger \beta g \neq 0$.

Suppose $\phi$ is a solution of \eqref{multi12}. Since, at any $(x^{4M},y^{4N})$ and for any $i,j\in\{1,\ldots,M\}$ with $i\neq j$,
\be
\frac{\partial^2 \phi}{\partial x_i^0 \partial x_j^0} = \frac{\partial^2 \phi}{\partial x_j^0 \partial x_i^0}\,,
\ee
we have that
\be
\frac{\partial}{\partial x_i^0} H_{x_j}\phi = \frac{\partial}{\partial x_j^0} H_{x_i} \phi
\ee
with $H_{x_j}\phi$ the right-hand side of \eqref{multi1}, where $H_{x_j}$ should \emph{not} be thought of as an operator on a Hilbert space, but rather as an operator acting on functions such as $\phi$, defined on $\sS_{xy}$ or $\Gamma(\RRR^4)^2$. Writing
\be
\frac{\partial}{\partial x_i^0} H_{x_j}\phi 
= H_{x_j} \frac{\partial\phi}{\partial x_i^0} + \Bigl[\frac{\partial}{\partial x_i^0}, H_{x_j} \Bigr] \phi\,, 
\ee
it follows further that
\be
\biggl([H_{x_j},H_{x_i}] +\Bigl[ i\frac{\partial}{\partial x_i^0} , H_{x_j} \Bigr] -\Bigl[ i\frac{\partial}{\partial x_j^0} , H_{x_i} \Bigr]  \biggr) \phi=0\,,
\ee
or, more compactly,\footnote{Alternatively, \eqref{xxconsistencyphi} can be obtained by the following reasoning. Let $K_j=i\partial/\partial x_j^0 -H_{x_j}$. If $\phi$ is a solution of the multi-time equations, then $K_j\phi=0$. Since $K_i$ is linear, also $K_iK_j\phi=0$. Likewise, $K_i\phi=0$ and $K_jK_i\phi=0$. Thus, $(K_jK_i-K_iK_j)\phi=0$.}
\be\label{xxconsistencyphi}
\biggl[ i\frac{\partial}{\partial x_i^0} -H_{x_i}, i\frac{\partial}{\partial x_j^0} - H_{x_j} \biggr] \phi =0\,.
\ee
For equations of the type of \eqref{multi12}, the last commutator does not involve any time derivatives, so we can think of $\phi$ as playing the role of an initial datum and therefore as arbitrary (cf.~\cite{pt:2013a} for more detail). Thus, consistency requires that
\be\label{xxconsistency}
\biggl[ i\frac{\partial}{\partial x_i^0} -H_{x_i}, i\frac{\partial}{\partial x_j^0} - H_{x_j} \biggr] =0\,.
\ee
Note that this condition means that the commutator vanishes on all functions, not merely on solutions of the multi-time equations \eqref{multi12}.
The conjunction of this condition and its sister conditions, i.e., the corresponding relations for $y_k$ and $y_\ell$ ($k,\ell\in\{1,\ldots,N\}$ with $k\neq \ell$) and for $x_j$ and $y_k$, will be called \emph{the consistency condition} in the following. The consistency condition forms the obvious analog of the condition \eqref{MTconsistency}, known \cite{pt:2013a} to characterize consistency in the case of a fixed number of particles. It will turn out (see below) that the consistency condition is actually necessary and sufficient for the consistency of multi-time equations; up to this point, we have only seen that it is necessary. 

Due to the nature of the operators $H_{x_j}$ as differential, multiplication, and insertion operators, the commutator in \eqref{xxconsistency} can actually be defined pointwise, so that \eqref{xxconsistency} can be satisfied in some region of $\Gamma(\RRR^4)^2$ and violated in another. Due to the nature of the reasoning that led to \eqref{xxconsistency}, the multi-time system \eqref{multi12} cannot be consistent in any region in which the consistency condition is violated.

We now show that \eqref{xxconsistency} is violated for the concrete equations \eqref{multi12} at non-spacelike configurations if $m_y>0$ and $g^\dagger \beta g \neq 0$. A calculation shows that, at all configurations and for any $m_y\geq 0$ and $g\in\CCC^4$,
\be\label{xxcommutator}
\biggl[ i\frac{\partial}{\partial x_i^0} -H_{x_i}, i\frac{\partial}{\partial x_j^0} - H_{x_j} \biggr] 
= \sum_{s=1}^4 g^*_s \Bigl( G_s(x_i-x_j)-G_s(x_j-x_i)  \Bigr)\,.
\ee
It is known \cite[Thm.~1.2 on p.~15]{thaller:1992} that 
\be
G(t,\vx) = i\Bigl( i \frac{\partial}{\partial t} -i\valpha\cdot \nabla +\beta m_y\Bigr) g\, \Delta(t,\vx) 
\ee
with $\Delta$ a certain scalar-valued distribution on $\RRR^4$ (depending on $m_y$), given explicitly in (e.g.)~\cite{thaller:1992}; it is actually a continuous function on the timelike and on the spacelike vectors, and a Dirac delta distribution on the light cone. Here, we need only its properties 
\begin{align}
\Delta(-x)&=-\Delta(x)\label{Deltaodd}\\
\Delta (x) &=0 \quad \text{for spacelike }x\label{Delta0}\\
\Delta(x) &\neq 0 \quad \text{for almost all timelike $x$ if }m_y>0\,.\label{Deltanonzero}
\end{align}
(For $m_y=0$, $\Delta(x) = -\mathrm{sgn}(x^0)(2\pi)^{-1} \, \delta(x^\mu x_\mu)$.)
Since the gradient of an odd function is even (i.e., $f_\mu=\partial \Delta/\partial x^\mu$ obeys $f_\mu(-x)=f_\mu(x)$), we obtain from \eqref{Deltaodd} that
\be
G(x)-G(-x)= 2i\beta m_y g \Delta(x)\,,
\ee
so that \eqref{xxconsistency} amounts to 
\be
m_y \; g^\dagger \beta g\; \Delta(x_i-x_j)=0\,.
\ee
We read off that this relation is always satisfied if $m_y=0$ or $g^\dagger \beta g=0$, but otherwise only where $\Delta(x_i-x_j)=0$; by \eqref{Delta0}, and since $\Delta(0)=0$ by \eqref{Deltaodd}, this is the case for spacelike configurations, and by \eqref{Deltanonzero}\footnote{In fact, from the explicit form of $\Delta$ \cite{thaller:1992} it can be read off that $\Delta$ vanishes for timelike $x$ only on countably many hypersurfaces.} it is almost never the case for configurations in which two $x$-particles are not spacelike separated. Thus, unless $m_y=0$ or $g^\dagger \beta g=0$, the system \eqref{multi12} is inconsistent on any open subset of $\Gamma(\RRR^4)^2$ containing a configuration in which some $x$-particles are not spacelike separated, in particular on $\Gamma(\RRR^4)^2$ itself.

(As an alternative argument, we can argue as follows from \eqref{xxconsistencyphi} instead of \eqref{xxconsistency} that the only solution of \eqref{multi12} on $\Gamma(\RRR^4)^2$ vanishes identically. Since every solution $\phi$ satisfies \eqref{xxconsistencyphi} and since, according to \eqref{xxcommutator}, the relevant commutator is a multiplication operator, $\phi$ must vanish wherever the right-hand side of \eqref{xxcommutator} does not---which is, assuming $m_y>0$ and $g^\dagger \beta g\neq 0$, at almost all configurations with timelike $x_i-x_j$. Taking $\phi$ to be smooth, it must vanish at all configurations with timelike $x_i-x_j$. Thinking of the evolution in just the variable $x_j$, it seems plausible from \eqref{multi1} that this can happen only if $\phi$ vanishes also for spacelike $x_i-x_j$. Thus, $\phi$ must vanish everywhere.)

For later use, we also report what the sister conditions of \eqref{xxconsistency} assert for \eqref{multi12}: The condition between $y_k$ and $y_\ell$ is always satisfied, and the condition between $x_j$ and $y_k$ is equivalent to
\be
\Bigl(i\frac{\partial}{\partial y^0}-H_y^\free\Bigr) G(y)=0\,,
\ee
a condition that holds by virtue of our assumption \eqref{Gdef1} that $G$ obeys the Dirac equation.

\subsection{Remark on the Computation of Consistency Conditions}
\label{sec:confusion}

A certain issue may easily be confusing when computing the commutators appearing in consistency conditions such as \eqref{xxconsistency}, and that is whether $H_{x_j}$ refers to the $j$-th $x$-variable or to the variable called $x_j$: The latter is correct, as we explain in this subsection.

We begin by explaining the confusing issue. For ease of notation, let us consider a simpler set of multi-time equations, involving only $x$-particles:
\be\label{simplemulti}
i\partial_{x_j^0} \phi(x_1,\ldots,x_N) = (H_{x_j}\phi)(x_1,\ldots,x_N) = G(x_j) \,\phi\bigl(x^{4N} \setminus x_j\bigr) \,.
\ee
Let us compute, for example, $\Big([H_{x_1},H_{x_3}]\phi\Big)(x_1,x_2,x_3,x_4)$. Of course, it is
\be
G(x_1) (H_{x_3}\phi)(x_2,x_3,x_4) - G(x_3)(H_{x_1}\phi)(x_1,x_2,x_4)\,,
\ee
but does $(H_{x_3}\phi)(x_2,x_3,x_4)$ mean $G(x_4)\, \phi(x_2,x_3)$ or $G(x_3)\, \phi(x_2,x_4)$? Or, as we put it above, does $H_{x_3}$ refer to the third variable in $(x_2,x_3,x_4)$, or to the one named $x_3$? The latter is correct.

To see this, it may be helpful to proceed step by step through the following consistency proof for \eqref{simplemulti}. Suppose we have a solution $\phi$ of \eqref{simplemulti}. Then 
\be\label{simplecon1}
i^2 \frac{\partial^2 \phi^{(4)}}{\partial x_1^0 \partial x_3^0}(x_1,x_2,x_3,x_4) = i^2 \frac{\partial^2\phi^{(4)}}{\partial x_3^0\partial x_1^0}(x_1,x_2,x_3,x_4)\,,
\ee
where the superscript ``(4)'' indicates that it is the 4-particle sector of $\phi$ that is being used.
By \eqref{simplemulti},
\be\label{simplecon2}
i\frac{\partial}{\partial x_1^0} \Bigl( G(x_3) \phi^{(3)}(x_1,x_2,x_4)\Bigr) = i\frac{\partial}{\partial x_3^0} \Bigl( G(x_1) \phi^{(3)}(x_2,x_3,x_4)\Bigr)\,.
\ee
Here, it is important to realize that $\partial/\partial x_3^0$ on the right-hand side means the derivative with respect to the variable called $x_3^0$, and not with respect to the third argument of $\phi^{(3)}$. That is because it initially meant the $x_3^0$-derivative of $i(\partial \phi^{(3)}/\partial x_1^0)(x_1,x_2,x_3,x_4)$ (where the question does not come up because the third variable is called $x_3$), and this expression could be rewritten in terms of $\phi^{(3)}$, shuffling around some variables. Thus, \eqref{simplecon2} is equivalent to
\be\label{simplecon3}
G(x_3) i \frac{\partial}{\partial x_1^0} \phi^{(3)}(x_1,x_2,x_4) = G(x_1) i \frac{\partial }{\partial x_3^0} \phi^{(3)}(x_2,x_3,x_4)
\ee
(still with $\partial/\partial x_3^0$ referring to the variable called $x_3$), which, by \eqref{simplemulti} again, is equivalent to
\be\label{simplecon4}
G(x_3)G(x_1) \phi(x_2,x_4) = G(x_1) G(x_3) \phi(x_2,x_4)\,.
\ee
At this point, no question of interpretation comes up any more. This equation is true, meaning that \eqref{simplemulti} satisfies the consistency condition.

\subsection{Consistency on Non-Spacelike Configurations}
\label{sec:Gamma}

Since we believe that physically reasonable examples of multi-time equations with particle creation and annihilation should be expected to be consistent \emph{only on spacelike} configurations, we regard the situation that the concrete system \eqref{multi12} has special cases ($m_y=0$ or $g^\dagger \beta g=0$) in which it is consistent also on non-spacelike configurations as a mere mathematical curiosity. Nevertheless, we give the argument for these special cases first because the argument is simpler than in the generic case. We have already seen that the consistency condition is satisfied on all of $\Gamma(\RRR^4)^2$; we now show that this is sufficient for consistency.

We write $\Gamma=\Gamma(\RRR^4)^2$ for short, and, for any $q^4=(x^{4M},y^{4N})\in\Gamma$,
\begin{subequations}
\begin{align}
\label{Kxdef}
K_{x_j}&= i\frac{\partial}{\partial x_j^0} - H_{x_j} \,,\\
\label{Kydef}
K_{y_k}&= i\frac{\partial}{\partial y_k^0} - H_{y_k} \,,
\end{align}
\end{subequations}
so that the multi-time equations \eqref{multi12} can be written as
\be\label{multi12K}
K_{x_j}\phi=0\,,\quad K_{y_k}\phi=0\,.
\ee
For any $q^4\in\Gamma$, let $L(q^4)$ denote the number of different values of time variables that occur in $q^4$; for example, if all time variables in $q^4$ have the same value, then $L(q^4)=1$. Let $\Gamma_L=\{q^4\in\Gamma:L(q^4)\leq L\}$; for example, $\psi$ can be regarded as defined on $\Gamma_1$. 
A configuration $q^4\in\Gamma_{L}\setminus \Gamma_{L-1}$ describes $L$ families of particles, each with a common time, say $q^4_\alpha=(x_\alpha^{4M_\alpha},y_\alpha^{4N_\alpha})$ ($\alpha=1,\ldots,L$) with $x^0_{\alpha,i}=x^0_{\alpha,j}=y^0_{\alpha,k}=y^0_{\alpha,\ell}=t_\alpha$; we also write $q^4_\alpha=(t_\alpha,q^3_\alpha)$ with $q^3_\alpha=(x^{3M_\alpha}_\alpha,y^{3N_\alpha}_\alpha)$ the spatial variables. We write $q^4$ either in the usual form $(x^{4M},y^{4N})$ or in the form $(t_1,q_1^3,\ldots,t_{L},q^3_{L})$. 
Since, at $q^4\in\Gamma_L\setminus\Gamma_{L-1}$,
\be
\frac{\partial \phi}{\partial t_\alpha} = \sum_{j=1}^{M_\alpha} \frac{\partial \phi}{\partial x_{\alpha,j}^0} + \sum_{k=1}^{N_\alpha} \frac{\partial\phi}{\partial y_{\alpha,k}^0}
\ee
and 
\be
K_\alpha \phi=0
\ee
for
\be\label{Kalphadef}
K_\alpha:=\sum_{j=1}^{M_\alpha} K_{x_{\alpha,j}} + \sum_{k=1}^{N_\alpha} K_{y_{\alpha,k}} \,,
\ee
any solution $\phi$ of \eqref{multi12} satisfies
\begin{align}\label{phiGamma1}
i\frac{\partial \phi}{\partial t_{\alpha}} 
&= \sum_{j=1}^{M_{\alpha}} H_{x_{\alpha,j}}\phi + \sum_{k=1}^{N_{\alpha}} H_{y_{\alpha,k}} \phi\\
&= \sum_{j=1}^{M_{\alpha}} \biggl\{ H^\free_{x_{\alpha,j}} \phi 
+ \sqrt{N+1}  \sum_{s_{N+1}=1}^4 g_{s_{N+1}}^* \, \phi_{s_{N+1}}\bigl(x^{4M}, (y^{4N}, x_{\alpha,j})\bigr)\nonumber \\
& \quad + \frac{1}{\sqrt{N}} \sum_{k=1}^N G_{s_k}(y_k - x_{\alpha,j}) \, \phi_{\widehat{s_k}}\bigl(x^{4M}, y^{4N} \backslash y_k\bigr) \biggr\}
+\sum_{k=1}^{N_{\alpha}} H^\free_{y_{\alpha,k}}\phi
\label{phiGamma2} 
\end{align}
for all $\alpha=1,\ldots,L$. Conversely, if \eqref{phiGamma2} holds at all $q^4\in\Gamma$ (and if $\phi$ is smooth on $\Gamma$) then also \eqref{multi12} holds.

To construct a solution $\phi$ from initial data $\phi_0$ at time 0, we proceed inductively along $L$: we first solve \eqref{phiGamma2} for $L=1$, then for $L=2$ etc.. To solve \eqref{phiGamma2} for $L=1$ means to determine $\psi$, i.e., to solve \eqref{Schr1Hdef}, which we take to be possible, ignoring the UV divergence. Suppose now that $\phi$ has been found on $\Gamma_{L-1}$; we will now construct it on $\Gamma_{L}$ in such a way that, with respect to the family times $t_\alpha$, $\phi$ satisfies the multi-time equations \eqref{phiGamma2}. We order the families so that $t_1\leq t_2 \leq \ldots \leq t_{L}$. To construct $\phi$ on $\Gamma_{L}$, we solve \eqref{phiGamma2} for $\alpha=L$ starting from initial data given by $\phi$ for $t_{L}=t_{L-1}$, treating $t_1,\ldots,t_{L-1}$ as fixed parameters. Note that these initial data are already defined because we assumed $\phi$ to be already defined on $\Gamma_{L-1}$. Note also that the $N+1$- and $N-1$-particle sectors of $\phi$ are always defined when needed because all arguments have time variables $t_1,\ldots,t_{L-1}$, or $t_{L}$.

We write $\Phi_L$ for the function thus constructed on $\Gamma_L$ from the function $\Phi_{L-1}$ given as $\phi$ on $\Gamma_{L-1}$. 
We need to verify that $\Phi_L$ satisfies \eqref{phiGamma2} for all $\alpha=1,\ldots,L$. For $\alpha=L$ that is clear because $\Phi_L$ was constructed as the solution of that equation. We now turn to $\alpha< L$. We need to show that $\Phi_{L,\alpha} := K_\alpha \Phi_L$ (with $K_\alpha$ as in \eqref{Kalphadef}) vanishes identically on $\Gamma_L$. Since \eqref{xxconsistency} means that $[K_{x_i},K_{x_j}]=0$, and likewise $[K_{y_k},K_{y_\ell}]=0$ and $[K_{x_j},K_{y_k}]=0$, we have that
\begin{equation}\label{KalphaKbeta}
\bigl[ K_\alpha,K_\beta\bigr]=0 \quad \text{for } \alpha,\beta\in\{1,\ldots,L\}\,.
\end{equation}
For $\beta=L$, and since $K_L\Phi_L=0$ by construction, we have that $K_LK_\alpha\Phi_L=K_\alpha K_L\Phi_L=0$, i.e., that the function $\Phi_{L,\alpha}$ satisfies \eqref{phiGamma2} with $\alpha$ replaced by $L$, with initial datum on $\Gamma_{L-1}$ given by $\Phi_{L-1,\alpha}=K_\alpha\Phi_{L-1}$ (constructed in a previous round of the induction). For $\alpha<L-1$, this means
\begin{equation}\label{KalphaPhiL-1}
\Phi_{L,\alpha}\bigl(t_1,q^3_1,\ldots,t_{L-1},q^3_{L-1},t_{L-1},q^3_L\bigr)=
K_\alpha \Phi_{L-1}\bigl(t_1,q^3_1,\ldots,t_{L-1},(q^3_{L-1},q^3_L)\bigr)\,.
\end{equation}
For $\alpha=L-1$, we need to be careful when formulating the initial condition because, due to the merger of families $L-1$ and $L$, $K_{L-1}$ means something else for $\Phi_{L-1}$ than for $\Phi_L$, namely $(K_L+K_{L-1})\Phi_L(t_L=t_{L-1})=K_{L-1}\Phi_{L-1}$; however, since $K_L\Phi_L=0$ by construction, it still follows that $K_{L-1}\Phi_L(t_L=t_{L-1})=K_{L-1}\Phi_{L-1}$; thus, \eqref{KalphaPhiL-1} applies also to $\alpha=L-1$.
By the linearity of \eqref{phiGamma2}, it suffices to show that the initial datum \eqref{KalphaPhiL-1} vanishes identically; that is, it suffices to show that $K_\alpha\Phi_{L-1}=0$. If $\alpha=L-1$, this is immediate from the construction of $\Phi_{L-1}$. For $\alpha<L-1$, this can be taken as an induction assumption. (Put differently, for $\alpha<L-1$, we repeat the above reasoning to find that it suffices to show that $K_\alpha\Phi_{L-2}=0$. After $L-\alpha$ repetitions we are done.) 

This completes our reasoning to the effect that the consistency condition (i.e., \eqref{xxconsistency} and the sister conditions), if valid on all of $\Gamma$, is sufficient for consistency of \eqref{multi12} on all of $\Gamma$. We now turn to justifying the consistency of \eqref{multi12} on $\sS_{xy}$ in the generic case ($m_y>0$ and $g^\dagger \beta g\neq 0$).

\subsection{Domain of Dependence}
\label{sec:domain}

As a preparation, we need a basic fact about the domain of dependence in the 1-time version \eqref{Schr1Hdef} of the evolution. To begin with, it is well known that, in the 1-particle Dirac equation, disturbances in the wave function propagate no faster than at the speed of light ($c=1$); that is, $\psi(t,\vx)$ is determined by the initial wave function $\psi(0,\cdot)$ on $\overline B_{|t|}(\vx)$, where
\be
\overline B_r(\vx) = \bigl\{\vy\in\RRR^3:\|\vy-\vx\|\leq r\bigr\}
\ee
is the closed 3-ball around $\vx$ of radius $r\geq 0$. One says that $\{0\}\times\overline B_{|t|}(\vx)$ is the \emph{domain of dependence} of $(t,\vx)$ at time 0. For any fixed number $N$ of particles, the 1-time Dirac equation of $N$ particles (either non-interacting or interacting by a potential) has the corresponding property \cite{pt:2013a} that $\psi(t,\vx_1,\ldots,\vx_N)$ is determined by the initial wave function $\psi(0,\cdot)$ on $\overline B^{(N)}_{|t|}(\vx_1,\ldots,\vx_N) \subset \RRR^{3N}$, where
\be
\overline B^{(N)}_{r}(\vx_1,\ldots,\vx_N)=\overline B_{r}(\vx_1)\times \cdots \times \overline B_{r}(\vx_N)
\ee
is the product of $N$ 3-balls.
For deriving Assertion~\ref{ass:consistency}, it plays a role to know something about the domain of dependence $M_{|t|}(x^{3M},y^{3N})$ of any configuration $(t,x^{3M},y^{3N})\in\RRR\times \Gamma(\RRR^3)^2$. Since it is a quite complicated set, we will, instead of dealing with it directly, give an upper bound, i.e., we will consider a simpler set $N_{|t|}(x^{3M},y^{3N})$ containing $M_{|t|}(x^{3M},y^{3N})$.

Since we are considering identical $x$-particles and identical $y$-particles, any two configurations differing only by a permutation of the $x$-particles and a permutation of the $y$-particles have the same $\psi$ up to a sign; so their $\psi$ values carry the same information, and the ordering of a configuration is irrelevant to the present purpose. For this reason, we will regard configurations in this subsection as \emph{unordered}. An unordered $x$-configuration can be regarded as a set $x^{3M}=\{\vx_1,\ldots,\vx_M\}$ or, when necessary, as a set-with-multiplicities represented by an occupation-number function $\nu:\RRR^3\to\{0,1,2,\ldots\}$ with $M=\sum_{\vx\in\RRR^3}\nu(\vx)<\infty$; the notation $x^{3M}\cup x^{\prime\, 3M'}$ then means $\nu+\nu'$. Likewise with $y$-configurations.

The domain of dependence can be characterized in the following way (that has a Bohmian or path-integral flavor). Imagine particles with trajectories starting in the configuration $(x^{3M},y^{3N})$, suppose that each particle can move at most at the speed of light, and suppose that $x$-particles can emit and absorb $y$-particles in the literal sense (i.e., $y$-world lines can begin and end on $x$-world lines). Then, for $t\geq 0$, $M_t(x^{3M},y^{3N})$ consists of all configurations that can be reached by our particles in a time interval of length $t$. Indeed, this is rather obvious from the Hamiltonian \eqref{Hdef}, in view of the facts that the free Hamiltonians allow the wave function to propagate at the speed of light for every particle variable, and the creation and annihilation terms allow $\psi$ at any configuration to influence $\psi$ at the configuration with one more $y$-particle, created at the location of an $x$-particle, and at the configuration with one $y$-particle removed, provided it was at the location of an $x$-particle.

To illustrate that $M_t(x^{3M},y^{3N})$ can be a complicated set, we note that it appears to be a non-trivial problem to decide for $M=1$ whether the $x$-particle can absorb all $y$-particles in a given configuration within time $t$.

The bigger set $N_t(x^{3M},y^{3N})$ is defined as follows, for $t\geq 0$. It contains all configurations that can be reached by the particles with trajectories as described above but with the modified rule that $y$-particles can disappear also when they do not collide with an $x$-particle. That is, it contains all configurations in which some $y$-particles have been removed, each remaining $y$-particle has traveled at most distance $t$, each $x$-particle has traveled at most distance $t$, and any number of further $y$-particles have been added, each within distance $t$ of the initial location of an $x$-particle. In formulas,\footnote{The notation in this formula is a bit sloppy in that it sometimes treats configurations as ordered and sometimes as unordered. This should not create any difficulty.}
\be\label{def_N_t}
\begin{array}{lll}
N_t(x^{3M},y^{3N}) = \Big\{ (X^{3M},Y^{3K}):&\\[3mm] 
\qquad 1.~ X^{3M} \in \overline B_{t}^{(M)}(x^{3M}) &\\[3mm] 
\qquad  2.~ Y^{3K} = \tilde y \cup \bigcup_{j=1}^M \hat{y}_j ~~\text{with} 
  &(i)~ \tilde y \in \overline B_{t}^{(L)}(\tilde{Y}^{3L}) ~\text{for}~ \tilde{Y}^{3L} \subset y^{3N} \\[3mm]
&(ii)~ \hat{y}_j \subset \overline B_{t}(\vx_j)
\Big\}.
\end{array}
\ee
From the path-characterization of $M_t(x^{3M},y^{3N})$ it is clear that
\be
N_t(x^{3M},y^{3N})\supset M_t(x^{3M},y^{3N})\,.
\ee
We have thus derived:

\begin{ass}
\label{ass:dependence}
Let $\psi$ be a solution of Equations \eqref{Schr1Hdef} with initial data $\psi(0,\cdot)$.
Then the initial data on $N_{|t|}(x^{3M},y^{3N})$ 
uniquely determine $\psi(t,x^{3M},y^{3N})$. 
\end{ass}

For a variant of the Hamiltonian \eqref{Hdef} with a UV cut-off, the corresponding statement is proven in \cite{pt:2013b}. 

\bigskip

\noindent\textbf{Remark.}
\begin{enumerate}
\setcounter{enumi}{\theremarks}
\item\label{rem:spacelikeNt} If the configuration $(t_1,q_1;t_2,q_2)$ is spacelike, and $t_2>t_1$, then also $(t_1,q_1;t_2',q_2')$ is spacelike for any $t_2'\in[t_1,t_2]$ and any $q_2'\in N_{t_2-t_2'}(q_2)$. This can be checked from the definition \eqref{def_N_t}, or is easy to see from the path-characterization of $N_t(q^3)$.
\end{enumerate}
\setcounter{remarks}{\theenumi}

\subsection{Consistency on Spacelike Configurations}
\label{sec:spacelike}

We will now complete the derivation of Assertion~\ref{ass:consistency} by showing that the consistency condition on $\sS_{xy}$ is sufficient for the consistency of \eqref{multi12} on $\sS_{xy}$.  
Let $L(q^4)$ denote again the number of different values of the time variables that occur in $q^4$, and let $\sS_L=\{q^4\in\sS_{xy}:L(q^4)\leq L\}$. We use again the notation of Section~\ref{sec:Gamma}, and note that \eqref{multi12K}--\eqref{phiGamma2} at $q^4\in\sS_L\setminus \sS_{L-1}$ are still valid for solutions of \eqref{multi12}. In particular, while some derivatives do not make immediate sense at some points in $\sS_{xy}$, as discussed in Remark~\ref{rem:tip} in Section~\ref{sec:MT-em-ab}, $\partial \phi/\partial t_\alpha$ does (and thus, all derivatives involved in $K_\alpha$ do) everywhere in $\sS_L\setminus\sS_{L-1}$. Moreover, according to what was laid down in Remark~\ref{rem:tip}, \eqref{multi12} implies that $K_\alpha\phi=0$ for all $\alpha$ everywhere in $\sS_L\setminus \sS_{L-1}$.

As in Section~\ref{sec:Gamma}, we assume that an initial datum $\phi_0$ is given and proceed inductively along $L$. To solve \eqref{phiGamma2} for $L=1$ means to determine $\psi$, i.e., to solve \eqref{Schr1Hdef}. Suppose now that $\phi$ has been found already on $\sS_{L-1}$; we will now construct it on $\sS_L$, obeying \eqref{phiGamma2}. Fix a $q^4\in\sS_L \setminus \sS_{L-1}$; we write $q^4=(t_1,q_1,\ldots,t_L,q_L)$ and order the families so that $t_1\leq t_2\leq \ldots \leq t_L$. To construct $\phi$ at $q^4$, we solve \eqref{phiGamma2} for $\alpha=L$ starting from initial data given by $\phi$ for $t_L=t_{L-1}$ (which are already defined by induction assumption), treating $t_1,\ldots,t_{L-1}$ as fixed parameters. 

Of course, once we fix $t_1,\ldots,t_{L-1}$, not all choices of spatial coordinates $q'_1,\ldots,q'_L$ will be such that
\be\label{tL-1q'}
\bigl(t_1,q'_1;\ldots;t_{L-1},q'_{L-1};t_{L-1},q'_L\bigr)
\ee
is a spacelike configuration. We now show that all configurations \eqref{tL-1q'} that we actually need are spacelike. Let $W(q^3)$ denote the set of configurations in $\Gamma(\RRR^3)^2$ obtained from $q^3\in\Gamma(\RRR^3)^2$ by erasing any of the $y$-particles; note that $W(x^{3M},y^{3N})$ has $2^N$ elements.  Let $n$ be the number of $y$-particles in $q_1,\ldots,q_{L-1}$. We claim that \eqref{phiGamma2} with $\alpha=L$ can be solved uniquely up to $q^4$ from initial data given only on the set
\be\label{L-1set}
\Bigl\{ (t_1,q_1';\ldots;t_{L-1},q'_{L-1};t_{L-1},q'_L): q'_1\in W(q_1),\ldots,q'_{L-1}\in W(q_{L_1}), q'_L\in N_{\Delta t}(q_L) \Bigr\}
\ee
with $\Delta t = t_L-t_{L-1}\geq 0$. Indeed, treating $q'_1,\ldots,q'_{L-1}$ as fixed parameters, and keeping in mind that there are $2^{n}$ choices of them, \eqref{phiGamma2} with $\alpha=L$ can be regarded as 1-time equations for $2^{n}$ wave functions $\psi(t'_L,q'_L)=\phi(t'_1,q'_1;\ldots;t'_L,q'_L)$ of the type \eqref{Schr1Hdef} plus coupling terms (viz., the terms involving $\phi(x^{4M},y^{4N}\setminus y_k)$ for $y_k$ in the $\alpha$-th family with $\alpha<L$). Without the coupling terms, Assertion~\ref{ass:dependence} would tell us that $N_{\Delta t}(q_L)$ is the domain of dependence for each of the $2^{n}$ wave functions; since they all have the same domain of dependence, this is also true in the presence of the coupling terms. Therefore, data on the set \eqref{L-1set} suffice for determining $\phi(q^4)$. By Remark~\ref{rem:spacelikeNt} in Section~\ref{sec:domain}, the set \eqref{L-1set} is contained in $\sS_{L-1}$, where $\phi$ is already defined. Thus, we have constructed $\phi$ on $\sS_L$.

To see that the $\phi$ thus constructed on $\sS_L$ satisfies \eqref{phiGamma2} for all $\alpha$ (not just $\alpha=L$), we can argue as in Section~\ref{sec:Gamma} in the paragraph containing \eqref{KalphaKbeta} and \eqref{KalphaPhiL-1}, using that the consistency condition holds on $\sS_{xy}$.\footnote{In fact, a weaker condition is sufficient: We only need that for any two particles, $j$ and $k$, the consistency condition for these two particles, $[K_j,K_k]=0$, holds on those spacelike configurations at which $j$ and $k$ do not collide. This fact is relevant for other examples of multi-time equations \cite{pt:2013d}, whose consistency conditions fail at such collisions. To see that this weaker condition is sufficient, note that it implies that $[K_\alpha,K_\beta]=0$ at every spacelike configuration at which no particle from family $\alpha$ collides with any particle from family $\beta$; let us call such configurations $\alpha$-$\beta$-safe. Since at $\alpha$-$\beta$-unsafe configurations, the $t_\beta$ variable cannot be increased independently of $t_\alpha$ without leaving the spacelike configurations (analogously to the situation of Remark~\ref{rem:tip} in Section~\ref{sec:MT-em-ab}), the relation $[K_\alpha,K_\beta]=0$ is not needed there. More precisely, we can argue as in the paragraph containing \eqref{KalphaPhiL-1} because, for any spacelike configuration $q^4\in\Gamma_L\setminus\Gamma_{L-1}$, its domain of dependence on configurations with $t_L=t_{L-1}$ and the ``cone'' in between do not contain any $L$-$\alpha$-unsafe configurations, so that $K_L\Phi_{L,\alpha}=0$ on that cone and, as a consequence, $\Phi_{L,\alpha}$ vanishes on $\Gamma_L\cap \sS_{xy}$. 

In fact, the necessary and sufficient condition for consistency is that $[K_\alpha,K_\beta]=0$ at every $\alpha$-$\beta$-safe configuration, or, equivalently: For every spacelike configuration $q=(q_1,\ldots,q_n)$ (with $q_j\in\RRR^4$), if particles $j$ and $k$ do not collide (i.e., $q_j\neq q_k$), then $[K_{C(j)},K_{C(k)}]=0$, where $K_{C(j)}$ is the sum of the $K_{j'}$ over all particles $j'$ colliding with $j$ (i.e., $q_{j'}=q_j$), including $j'=j$.}

We have thus constructed a $\phi$ on $\sS_{xy}$ that solves \eqref{phiGamma2} everywhere and therefore \eqref{multi12}. The construction also makes clear that the solution of \eqref{multi12} on $\sS_{xy}$ is unique. This completes the derivation of Assertion~\ref{ass:consistency}.

\bigskip

As a by-product of this derivation, we obtain further that also in the multi-time evolution \eqref{multi12}, disturbances in the wave function propagate no faster than at the speed of light. More precisely, let $J^+(x)$ denote the closed future light cone of $x$ and let, for any space-time region $R$, $J^+(R)=\cup_{x\in R}J^+(x)$ denote its causal future and $J^-(R)$ its causal past. 

\begin{ass}\label{ass:finite_speed2}
Consider the multi-time emission--absorption model \eqref{multi12} with two different initial conditions $\phi_0$ and $\phi'_0$ at time $0$ that differ only on configurations with at least one particle in the region $R\subset \{(t,\vx)\in\RRR^4:t=0\}$. Then
\be
\phi'(x_1,\ldots,x_M, y_1,\ldots,y_N) = \phi(x_1,\ldots,x_M, y_1,\ldots,y_N)
\ee
whenever all $x_1,\ldots,x_M, y_1,\ldots,y_N \notin J(R)=J^+(R) \cup J^-(R)$.
\end{ass}

\begin{ass}\label{ass:finite_speed1}
Consider the multi-time emission--absorption model \eqref{multi12} with additional external fields $A_{\mu}$ acting on the $x$ and $y$ particles; i.e., add $\sum_{a=1}^3(\alpha_a)_{r_jr'_j}e_xA_a(x_j)+e_xA_0(x_j)\delta_{r_jr'_j}$ to the bracket in \eqref{Hxjfreedef}, and correspondingly in \eqref{Hykfreedef}, with real constants $e_x$ and $e_y$ (the charges of the $x$ and $y$ particles). Consider two choices $A_{\mu}$ and $A'_{\mu}$ of the external field that differ only in a space-time region $R \subset \{(t,\vx)\in\RRR^4:t>0\}$, so for any given initial condition at $t=0$ we consider two multi-time wave functions $\phi$ and $\phi'$, the solutions of the multi-time equations with $A_{\mu}$ and $A'_{\mu}$. Then
\be
\phi'(x_1,\ldots,x_M, y_1,\ldots,y_N) = \phi(x_1,\ldots,x_M, y_1,\ldots,y_N)
\ee
whenever all $x_1,\ldots,x_M, y_1,\ldots,y_N \notin J^+(R)$.
\end{ass}

\subsection{Tomonaga--Schwinger Theory}
\label{proofs:Tomonaga-Schwinger}

We will turn to Assertion~\ref{ass:unitary_HS_evolution} later and first deal with Assertion~\ref{ass:Tomonaga-Schwinger}.


\begin{proof}[Derivation of Assertion~\ref{ass:Tomonaga-Schwinger}]
We first compute $\Hd_I(x)$. The Hamiltonian corresponding to the multi-time equations \eqref{multi12} is given in \eqref{Hdef}. The interaction Hamiltonian is thus
\begin{align}
H_I 
&= \sum_{i=1}^N \sum_{s=1}^4 \Bigl[ g_s^* \, b_s(\vx_i) + g_s \, b^\dagger_s(\vx_i)\Bigr]\\
&= \sum_{r,s=1}^4 \int d^3\vx \Bigl[  a^\dagger_r(\vx) \,g_s^*\, b_s(\vx)\, a_r(\vx) + a^\dagger_r(\vx) \, g_s\, b^\dagger_s(\vx) \, a_r(\vx) \Bigr]\,,\label{TSHI2}
\end{align}
where $a_r(\vx)$ is the annihilation operator for an $x$-particle at location $\vx$ with spinor $e_r$ (with the $e$'s the standard basis in $\CCC^4$), and $b_s(\vy)$ correspondingly for $y$-particles. Now consider the interaction picture (in the non-relativistic case, not considering any curved hypersurfaces in space-time). The Hamiltonian that occurs in the Schr\"odinger equation in the interaction picture is
\be
H_I^\ip (t) = \sum_{r,s=1}^4 \int d^3\vx \, e^{iH_\free t} \Bigl[ \cdots \Bigr] e^{-iH_\free t}
\ee
with $[\cdots]$ the same expression as in square brackets in \eqref{TSHI2}. Since, on the other hand, $H_I^\ip(t) = \int d^3\vx \, \Hd_I(t,\vx)$, we read off that
\begin{align}
\Hd_I(t,\vx)
&= \sum_{r,s=1}^4 e^{iH_\free t} \Bigl[ \cdots \Bigr] e^{-iH_\free t}\\
&= F_{\Sigma_t\to \Sigma_0} \Bigl( N_x(\vx) \otimes \sum_{s=1}^4 \bigl( g_s^*\, b_s(\vx)+g_s \, b^\dagger_s(\vx) \bigr) \Bigr) F_{\Sigma_0\to\Sigma_t}\,,
\label{HdIdef}
\end{align}
where $F_{\Sigma\to\Sigma'}$ is the free time evolution, here $F_{\Sigma_0\to\Sigma_t}=e^{-iH_\free t}$.
We have thus obtained the expression \eqref{HdIdef2}=\eqref{HdIdef} for $\Hd_I(x)$, $x\in \RRR^4$. 

We now compute what the multi-time equations \eqref{multi12} look like in a different Lorentz frame. In this regard, note that \eqref{multi12} are \emph{not} Lorentz invariant, so the transformed equation should not agree with the original one. In addition, we will use one convention differently than usual: Usually, when transforming (e.g.)\ the free Dirac equation, one also applies a transformation matrix in spin space; in other words, along with a change of basis in space-time, one also changes basis in spin space; we will not do that. We will keep the basis in spin space fixed once and for all; this choice will make later parts of the reasoning more transparent. With this convention, it follows that also the free Dirac equation changes form under Lorentz transformations, viz., the gamma matrices have to be replaced by modified matrices. So let us consider, apart from the Lorentz frame $L$ in which \eqref{multi12} holds, another  Lorentz frame $\underline{L}$. We write $u^\mu$ for the components of the vector $u$ with respect to the basis $L$ and $u^{\underline{\mu}}$ for the components of $u$ with respect to the basis $\underline{L}$; thus, in our notation, $u^{\underline{0}}$ is a different number than $u^0$. Let $\Lambda$ be the transformation matrix,
\be\label{Lambdadef}
u^\mu = \Lambda^\mu_{\:\:\underline{\nu}} u^{\underline{\nu}}\,.
\ee
It follows that
\be\label{Lambdadphi}
\partial_{x^\mu_j} \phi = \Lambda_\mu^{\:\:\underline{\nu}} \partial_{x^{\underline{\nu}}_j} \phi\,,
\ee
where the indices of $\Lambda^\mu_{\:\:\underline{\nu}}$ get raised and lowered with $g_{\mu\nu}$ and $g^{\underline{\mu}\underline{\nu}}$, and $\partial_{x^{\underline{\nu}}_j} \phi$ means the directional derivative of $\phi$ in the direction of the $\underline{\nu}$-th basis vector of $\underline{L}$ (or, equivalently, the gradient of $\phi$ expanded in the basis $\underline{L}$); for simplicity, we think of $\phi$ as the same function on the space-time manifold $\sM$ and the spacelike set $\sS$ formed from it (rather than considering a very different coordinate expression in $\underline{L}$). In accordance with what we said above about the spin basis, we write $\gamma^{\underline{\nu}}$ for the four matrices satisfying
\be
\gamma^\mu = \Lambda^\mu_{\:\:\underline{\nu}} \gamma^{\underline{\nu}}\,.
\ee
Recall that \eqref{multi1} is of the form
\be\label{multi_time_form_x}
i\partial_{x^0_j}\phi = (\gamma^0_{j})^{-1}m_x\phi -i\sum_{a=1}^3(\gamma^0_{j})^{-1} \gamma^a_{j}\partial_{x_j^a} \phi + \Ann\phi + \Cr\phi\,, 
\ee
where we have written out $H^\free_{x_j}$, used the notation $\gamma_j^\mu$ again for $\gamma^\mu$ acting on $r_j$ (while $\gamma_k^\mu$ will act on $s_k$, so the letters $j$ and $k$ also indicate the particle species), and introduced the abbreviations $\Ann \phi$ and $\Cr \phi$ for the annihilation and creation terms in \eqref{multi1}. Since $(\gamma^0)^{-1}=\gamma^0$, we can simply write $\gamma^0$ instead of $(\gamma^0)^{-1}$; the same applies to $\gamma^\mu n_\mu$ for any timelike vector $n_\mu$, in particular to $\gamma^{\underline{0}}$. Expanding $\partial\phi$ in \eqref{multi_time_form_x} using \eqref{Lambdadphi} and sorting by components of $\partial_{x_j^{\tilde{\nu}}}\phi$, we obtain that
\begin{align}\label{Lambda_multi_time_equations_explicit_nocut_x}
i\partial_{x^{\underline{0}}_j}\phi 
&= \gamma^{\underline{0}}_{j} m_x\phi -i\sum_{\underline{a}=1}^3 \gamma^{\underline{0}}_{j} \gamma^{\underline{a}}_{j}\partial_{x_j^{\underline{a}}} \phi + \gamma^{\underline{0}}_{j} \gamma^0_{j} \Bigl[\Ann\phi + \Cr\phi\Bigr]\\
&=: \underline{H}^\free_{x_j} \phi + \gamma^{\underline{0}}_{j} \gamma^0_{j} \Bigl[\Ann\phi + \Cr\phi\Bigr]\,.
\label{Lambda_multi_time_equations_nocut_x}
\end{align}
Likewise, the transform of \eqref{multi2} for $y$-particles reads
\be\label{Lambda_multi_time_equations_nocut_z}
i\partial_{y^{\underline{0}}_k}\phi = \underline{H}^\free_{y_k} \phi \,.
\ee

We now turn to computing the time evolution of 
$\tilde\psi_\Sigma = \psi_\Sigma^\ip = F_{\Sigma\to \Sigma_0} \psi_\Sigma$. 
To this end, consider two infinitesimally neighboring spacelike hypersurfaces $\Sigma,\Sigma'$ (as would arise from a smooth mapping $f:\RRR^4\to\sM$ as $\Sigma=f(t,\RRR^3)$ and $\Sigma'=f(t+dt,\RRR^3)$), and let $n^\mu(x)$ denote the future-pointing unit normal vector field on $\Sigma$. Let $d\ell$ denote some positive infinitesimal number, and let $\tau(x) \, d\ell$ denote the signed thickness of the layer between $\Sigma$ and $\Sigma'$ at $x\in \Sigma$ (so $\tau(x)>0$ [respectively, $<0$] where $\Sigma'$ lies in the future [respectively, past] of $\Sigma$); that is, for any $x \in\Sigma$,
\be
(x')^\mu := x^\mu + \tau(x) \, n^\mu(x) \, d\ell
\ee
lies on $\Sigma'$. Note that (to first order in $d\ell$)
\be\label{psiSigmaprime}
\psi_{\Sigma'}\bigl( x^{\prime 4M}, y^{\prime 4N}\bigr) = \psi_\Sigma(x^{4M},y^{4N}) + d\ell \sum_{j=1}^M \tau(x_j) \, n^\mu(x_j) \frac{\partial \phi}{\partial x^\mu_j} + d\ell\sum_{k=1}^N \tau(y_k) \, n^\mu(y_k) \frac{\partial \phi}{\partial y^\mu_k}\,.
\ee
Conversely, if $\phi^\free$ denotes the solution of the \emph{free} multi-time equations (i.e., with $\Ann\phi+\Cr\phi$ omitted) with initial condition $\psi_{\Sigma'}$ on $\Sigma'$, we obtain that
\begin{multline}\label{FpsiSigma}
F_{\Sigma' \to \Sigma}\psi_{\Sigma'}\bigl( x^{4M}, y^{4N}\bigr)\\ 
= \psi_{\Sigma'}(x^{\prime 4M},y^{\prime 4N}) - d\ell \sum_{j=1}^M \tau(x_j) \, n^\mu(x_j) \frac{\partial \phi^\free}{\partial x^\mu_j} - d\ell\sum_{k=1}^N \tau(y_k) \, n^\mu(y_k) \frac{\partial \phi^\free}{\partial y^\mu_k}\,.
\end{multline}
(It does not matter whether $\partial \phi^\free$ is evaluated at $(x^{\prime 4M},y^{\prime 4N})$ or at $(x^{4M},y^{4N})$ because the difference is of higher order in $d\ell$.) Now use \eqref{Lambda_multi_time_equations_nocut_x} for $\phi^\free$ without the terms $\Ann\phi+\Cr\phi$ in the Lorentz frame $\underline{L}$ that is tangent to $\Sigma$ at $x_j$ to find that
\be
n^\mu(x_j) \frac{\partial \phi^\free}{\partial x^\mu_j} = \partial_{x^{\underline{0}}_j}\phi^\free = 
-i\underline{H}^\free_{x_j} \phi^\free = -i\underline{H}^\free_{x_j} \psi_{\Sigma'}\,.
\ee
With the same reasoning for $y_k$, we have that
\begin{align}
F_{\Sigma' \to \Sigma}\psi_{\Sigma'}\bigl(x^{4M},y^{4N}\bigr) &= \psi_{\Sigma'}\bigl( x^{\prime 4M}, y^{\prime 4N}\bigr) - d\ell \sum_{j=1}^M \tau(x_j) \, \bigl( -i \underline{H}^{\free}_{x_j} \bigr) \phi \bigl( x^{4M}, y^{4N}\bigr) \nonumber \\
&\quad - d\ell \sum_{k=1}^N \tau(y_k) \, \bigl( -i \underline{H}^{\free}_{y_k} \bigr) \phi \bigl( x^{4M}, y^{4N}\bigr) \,,
\end{align}
where $\underline{L}$ is chosen differently (viz., tangent to $\Sigma$) at every $x_j$ and $y_k$. It follows that (to first order in $d\ell$)
\begin{align}
i\bigl(F_{\Sigma' \to \Sigma}\psi_{\Sigma'}-\psi_\Sigma\bigr) \bigl(x^{4M},y^{4N}\bigr) &=  d\ell  \sum_{j=1}^M \tau(x_j) \left( i\frac{\partial}{\partial x^{\underline{0}}_j} - \underline{H}^{\free}_{x_j} \right) \phi \bigl( x^{4M}, y^{4N}\bigr)  \nonumber \\
&+\: d\ell \sum_{k=1}^N \tau(y_k) \left( i\frac{\partial}{\partial y^{\underline{0}}_k} - \underline{H}^{\free}_{y_k} \right) \phi \bigl( x^{4M}, y^{4N}\bigr)
\end{align}
[using \eqref{Lambda_multi_time_equations_nocut_x} and \eqref{Lambda_multi_time_equations_nocut_z}]
\begin{align}
&= d\ell \sum_{j=1}^M \tau(x_j) \gamma^{\underline{0}}_{j}\gamma^0_{j} \Bigl[\Ann\phi + \Cr\phi\Bigr] \bigl( x^{4M}, y^{4N}\bigr)\\
&= d\ell \sum_{j=1}^M \tau(x_j) \gamma^{\underline{0}}_{j}\gamma^0_{j} 
\Bigg[ \sqrt{N+1}  \sum_{s_{N+1}=1}^4 g_{s_{N+1}}^* \, \psi_{\Sigma,s_{N+1}}\bigl(x^{4M}, (y^{4N}, x_j)\bigr)\nonumber \\
& \quad\quad + \frac{1}{\sqrt{N}} \sum_{k=1}^N G_{s_k}(y_k - x_j) \, \psi_{\Sigma,\widehat{s_k}}\bigl(x^{4M}, y^{4N} \backslash y_k\bigr)\Bigg] \label{psiSigmaevol}
\end{align}

We now put \eqref{HdIdef} and \eqref{psiSigmaevol} together and compare to the Tomonaga--Schwinger equation \eqref{TS}. The left-hand side of the Tomonaga--Schwinger equation is
\be\label{lhsTS}
\text{lhs}:=i\bigl(\tilde\psi_{\Sigma'}-\tilde\psi_\Sigma\bigr) = i\bigl(F_{\Sigma' \to \Sigma_0}\psi_{\Sigma'}-F_{\Sigma \to \Sigma_0}\psi_\Sigma\bigr)
=iF_{\Sigma \to \Sigma_0}\bigl(F_{\Sigma' \to \Sigma}\psi_{\Sigma'}-\psi_\Sigma\bigr)\,.
\ee
The right-hand side of the Tomonaga--Schwinger equation is, using \eqref{HdIdef},
\begin{align}
\text{rhs}&:= \biggl( \int_{\Sigma}^{\Sigma'} \!\!\!\! d^4 x \, \Hd_I(x)\biggr)\, \tilde\psi_\Sigma\\
&= d\ell\biggl( \int_{\Sigma} d^3 x \, \tau(x) \Hd_I(x)\biggr)\, F_{\Sigma \to \Sigma_0} \psi_\Sigma \\
&= d\ell\,F_{\Sigma \to \Sigma_0}\biggl( \int_{\Sigma} d^3 x \, \tau(x) F_{\Sigma_0 \to \Sigma} \,\Hd_I(x)\, F_{\Sigma \to \Sigma_0}\biggr) \psi_\Sigma\\
&= d\ell\,F_{\Sigma \to \Sigma_0}\biggl( \int_{\Sigma} d^3 x \, \tau(x) \,F_{\Sigma_{x^0}\to \Sigma} \Bigl[ N_x(\vx)\otimes \sum_{s=1}^4 \bigl( g_s^*\, b_s(\vx)+g_s \, b^\dagger_s(\vx) \bigr) \Bigr] F_{\Sigma\to\Sigma_{x^0}}\biggr) \psi_\Sigma\,.\label{rhsTS}
\end{align}
Let us write $t$ for $x^0$.
Since, under the free time evolution, $x$-particles and $y$-particles do not interact, we have that
\begin{multline}
F_{\Sigma_t\to \Sigma} \Bigl[ N_x(\vx)\otimes \sum_{s=1}^4 \bigl( g_s^*\, b_s(\vx)+g_s \, b^\dagger_s(\vx) \bigr) \Bigr] F_{\Sigma\to\Sigma_t}= \\
\Bigl( F^x_{\Sigma_t\to \Sigma}N_x(\vx)F^x_{\Sigma\to\Sigma_t}\Bigr)\otimes  \Bigl( F^y_{\Sigma_t\to \Sigma} \sum_{s=1}^4 \bigl( g_s^*\, b_s(\vx)+g_s \, b^\dagger_s(\vx) \bigr)  F^y_{\Sigma\to\Sigma_t}\Bigr)
\end{multline}
We abbreviate the last expression as $A(x)\otimes B(x)$ with $x=(t,\vx)$. Since both $F^x_{\Sigma_t\to \Sigma}$ and $N_x(\vx)$ leave each sector of the $x$-Fock space invariant, so does $A(x)$; on the sector with $M$ $x$-particles, $F^x_{\Sigma_t\to \Sigma}$ is the tensor product of $M$ 1-particle operators, while $N_x(\vx)$ is the sum of $M$ 1-particle operators as in \eqref{Ndeltapsi}. Thus,
\be
A(x)\Big|_{\Hilbert_\Sigma^{x,M}} = \sum_{j=1}^M F^{x_j}_{\Sigma_t\to\Sigma} \,\delta^3(\vx_j-\vx) \, F^{x_j}_{\Sigma\to\Sigma_t}
\ee
with $\delta^3(\vx_j-\vx)$ regarded as a multiplication operator on $L^2(\RRR^{3M},\CCC^k)$. From the transformation behavior of Green functions as described in the paragraph containing \eqref{Green1} and \eqref{Green2} in Section~\ref{sec:LorentzTrafo}, it follows that, for the free 1-particle Dirac equation (in the coordinates $t_1,\vx_1$) and a spacelike hypersurface $\Sigma$ passing through $(t,\vx)$,
\be\label{gammagammadelta}
F_{\Sigma_t\to\Sigma} \, \delta^3(\vx_1-\vx) \, F_{\Sigma\to\Sigma_t} = \gamma^{\underline{0}} \gamma^0\, \delta^3_\Sigma(\vx_1-\vx)
\ee
with $\underline{L}$ the Lorentz frame tangent to $\Sigma$ in $(t,\vx)$ and $\delta^3_\Sigma$ the delta function on $\Sigma$. For obtaining this relation, one should keep in mind our unusual convention about fixing the basis in spin space and note that the fact that $\Sigma$ is curved plays no role since the solution vanishes outside the light cone of $(t,\vx)$. We also note, since this is perhaps not obvious from the expression, that the right-hand side of \eqref{gammagammadelta} is actually self-adjoint as a multiplication operator on $\Hilbert_{1,\Sigma}$ as in \eqref{scpHilbert1Sigma}.\footnote{For verifying this, it is useful to note that $\gamma^{\underline{0}}\gamma^0=\gamma^\mu n_\mu(x) \gamma^0$ is a self-adjoint $4\times 4$ matrix and that $\gamma^{\underline{0}}$ is its own inverse as a $4\times 4$ matrix; the latter fact follows from the relation $\gamma^\mu\gamma^\nu + \gamma^\nu\gamma^\mu = 2g^{\mu\nu} I$ (with $I$ the identity matrix) by contracting with $n_\mu(x)n_\nu(x)$ and using that $n_\mu(x) n^\mu(x)=1$. \label{fn:gammaunderline0}}

We thus obtain that $A(x)$ acts as the multiplication operator
\be\label{Axgammadelta}
A(x)\Big|_{\Hilbert_\Sigma^{x,M}} = \sum_{j=1}^M \gamma^{\underline{0}}_{j} \gamma^0_{j} \,\delta^3_\Sigma(\vx_j-\vx)\,. 
\ee
It follows that
\be
\int_\Sigma d^3x\, \tau(x)\, A(x)\otimes B(x)\, \psi_\Sigma (x^{4M},y^{4N})=
\sum_{j=1}^M \tau(x_j) \, \gamma^{\underline{0}}_{j} \gamma^0_{j} B(x_j) \,\psi_\Sigma(x^{4M},y^{4N})\,.
\ee
To evaluate $B(x_j)\psi_\Sigma$, we may think in terms of the solution $\phi^{\free}$ of the \emph{free} multi-time equations with initial condition on $\Sigma$ given by $\psi_\Sigma$. Since the free time evolution acts on each sector (such as the $(N+1)$-$y$-particle sector) separately, and within each sector on each particle separately, one easily sees that
\begin{align}
B(x_j) \psi_\Sigma(x^{4M},y^{4N}) &= \sqrt{N+1}  \sum_{s_{N+1}=1}^4 g_{s_{N+1}}^* \, \psi_{\Sigma,s_{N+1}}\bigl(x^{4M}, (y^{4N}, x_j)\bigr)\nonumber \\
& \quad\quad + \frac{1}{\sqrt{N}} \sum_{k=1}^N G_{s_k}(y_k - x_j) \, \psi_{\Sigma,\widehat{s_k}}\bigl(x^{4M}, y^{4N} \backslash y_k\bigr)\,.
\end{align}

Putting this together with \eqref{psiSigmaevol}, \eqref{lhsTS}, and \eqref{rhsTS}, we obtain that lhs = rhs; that is, we obtain that the Tomonaga--Schwinger equation holds.
\end{proof}


As a preparation for deriving the further assertions of Section~\ref{sec:TS}, we need an auxiliary assertion. Let $a_r(x)$ denote the Heisenberg-evolved annihilation operator,
\be\label{atdef}
a_r(t,\vx) = e^{iHt}\, a_r(\vx)\, e^{-iHt}\,,
\ee
with $H$ the single-time Hamiltonian as in \eqref{Hdef}. Let $\Sigma$ be a spacelike hypersurface.
Recall that $\Hilbert_\Sigma$, as defined around \eqref{HilbertSigmadef}, is a space of functions $\psi_\Sigma(x_1\ldots x_M,y_1\ldots y_N)$ of arguments $x_j,y_k\in\Sigma$. We define $a_\Sigma,b_\Sigma$ as the ``literal'' or ``immediate'' annihilation operators on $\Hilbert_\Sigma$. That is, let $e_1\ldots e_4$ be the standard basis in spin space $\CCC^4$; for $x,y\in\Sigma$ and $r,s\in\{1\ldots 4\}$, we define the annihilation and creation operators
\begin{align}
a_{\Sigma,r}(x)\, \psi_{\Sigma,r_1\ldots r_M}(x^{4M},y^{4N}) 
&=\sqrt{M+1}\; (-1)^M\, \psi_{\Sigma,r_1\ldots r_M,r}\bigl((x^{4M},x),y^{4N} \bigr)\label{aSigmadef}\\
a^\dagger_{\Sigma,r}(x) \,\psi_{\Sigma,r_1\ldots r_M}(x^{4M},y^{4N})
&= \frac{1}{\sqrt{M}} \sum_{j=1}^M (-1)^{j+1} \, \bigl(\gamma^\mu n_\mu(x)\gamma^0\bigr)_{r_j r}\; \delta^3_\Sigma(x_j-x)\:\times \nonumber\\
& \qquad \times\: \psi_{\Sigma,\widehat{r_j}}(x^{4M}\setminus x_j,y^{4N})\\
b_{\Sigma,s}(y)\, \psi_{\Sigma,s_1\ldots s_N}(x^{4M},y^{4N})
&= \sqrt{N+1}\, \psi_{\Sigma,s_1\ldots s_N,s}\bigl(x^{4M},(y^{4N},y)\bigr)\\
b^\dagger_{\Sigma,s}(y) \,\psi_{\Sigma,s_1\ldots s_N}(x^{4M},y^{4N})
&= \frac{1}{\sqrt{N}} \sum_{k=1}^N \bigl(\gamma^\mu n_\mu(y)\gamma^0\bigr)_{s_k s}\; \delta^3_\Sigma(y_k-y)\:\times \nonumber\\
&\qquad \times\: \psi_{\Sigma,\widehat{s_k}}(x^{4M},y^{4N}\setminus y_k)\,,
\end{align}
where $n_\mu(x)$ is again the future-pointing unit normal vector to $\Sigma$ at $x$, and summation over $\mu$ is understood.
The hat in $\widehat{r_j}$ means omission. Not all spin indices are written out in these expressions.\footnote{For verifying that the expression given for $a^\dagger_\Sigma$ is indeed the adjoint in $\Hilbert_\Sigma$ of the expression given for $a_\Sigma$, one uses the facts mentioned in Footnote~\ref{fn:gammaunderline0}.}
Note that these ``literal'' annihilation and creation operators automatically satisfy the canonical (anti-)commutation relations (CAR/CCR):
\begin{align}
\label{CARCCRfirst}
\{a_{\Sigma,r}(x),a_{\Sigma,r'}(x')\}&=0\\
\{a^\dagger_{\Sigma,r}(x),a^\dagger_{\Sigma,r'}(x')\}&=0\\
\{a_{\Sigma,r}(x),a^\dagger_{\Sigma,r'}(x')\}&=\bigl( \gamma^\mu n_\mu(x) \gamma^0 \bigr)_{rr'}\;\delta^3_\Sigma(x-x')\label{CARaadagger}\\
[b_{\Sigma,s}(y),b_{\Sigma,s'}(y')]&=0\\
[b^\dagger_{\Sigma,s}(y),b^\dagger_{\Sigma,s'}(y')]&=0\\
[b_{\Sigma,s}(y),b^\dagger_{\Sigma,s'}(y')]&=\bigl(\gamma^\mu n_\mu(y) \gamma^0\bigr)_{ss'}\;\delta^3_\Sigma(y-y')\\
[a^{\#}_{\Sigma,r}(x),b^{\circ}_{\Sigma,s}(y)]&=0
\label{CARCCRlast}
\end{align}
for any $x,x',y,y'\in\Sigma$, where $a^{\#}$ means either $a$ or $a^\dagger$, and $b^\circ$ means either $b$ or $b^\dagger$.

\begin{ass}\label{ass:axfree}
Define
\be\label{afreetdef}
a_r^\free(t,\vx) = e^{iH^\free t}\, a_r(\vx)\, e^{-iH^\free t}\,.
\ee
Then, for any spacelike hypersurface $\Sigma$ and any $x\in\Sigma$,
\be\label{afreeaSigma}
a_{r}^\free(x)= F_{\Sigma\to\Sigma_0} a_{\Sigma,r}(x) F_{\Sigma_0\to\Sigma}\,.
\ee
As a consequence, for any two spacelike hypersurfaces $\Sigma,\hat\Sigma$ both containing $x$,
\be\label{aSigmaprimeSigmafree}
a_{\hat\Sigma,r}(x) =
F_{\Sigma\to\hat\Sigma} a_{\Sigma,r}(x) F_{\hat\Sigma\to\Sigma}\,.
\ee
Likewise for $b^\free_s(y)$ and $b_{\Sigma,s}(y)$.
\end{ass}

\begin{proof}[Derivation]
For the sake of this argument, let $a_r(\vx)$, $a_r^{\free}(x)$, and $a_{\Sigma,r}(x)$ be defined by \eqref{adef}, \eqref{afreetdef}, and \eqref{aSigmadef}, respectively, even for wave functions that are \emph{not necessarily} anti-symmetric in the $x$-variable or symmetric in the $y$-variables. Now consider the case that, for a particular choice of $M,N$, the $(M+1,N)$-particle sector of $\psi_0$ is a tensor product of some $(M,N)$-particle wave function $\chi$ and a 1-$x$-particle wave function $\varphi$ (applied to $x_{M+1}$). Since the free time evolution evolves each particle separately, we have that 
\be
a_r^{\free}(x) \psi_0(x^{3M},y^{3N}) = \varphi_r(x) \, \chi(x^{3M},y^{3N})
\ee
for any $(x^{3M},y^{3N})\in (\RRR^3)^{M+N}$ and
\be
a_{\Sigma,r}(x) \psi_\Sigma (x^{4M},y^{4N}) = \varphi_r(x) \, (F_{\Sigma_0\to\Sigma}\chi)(x^{4M},y^{4N})
\ee
for any $(x^{4M},y^{4N})\in\Sigma^{M+N}$. It follows that \eqref{afreeaSigma} holds for such a special $\psi_0$. By linearity, it holds for any $\psi_0$; in particular, it holds for $\psi_0$ that is anti-symmetric in the $x$-variables and symmetric in the $y$-variables.
\end{proof}

As a corollary of Assertion~\ref{ass:axfree}, \eqref{HdIdef} can be rewritten as
\be\label{HdIxSigma}
\Hd_I(x)= F_{\Sigma\to \Sigma_0} \Bigl( \sum_{r=1}^4 a^{\dagger}_{\Sigma,r}(x)a_{\Sigma,r}(x) \otimes \sum_{s=1}^4 \bigl( g_s^*\, b_{\Sigma,s}(x)+g_s \, b^\dagger_{\Sigma,s}(x) \bigr) \Bigr) F_{\Sigma_0\to\Sigma}
\ee
for any spacelike hypersurface $\Sigma$ containing $x$. (We note that $\sum_r a^\dagger_{\Sigma,r}(x) \, a_{\Sigma,r}(x)$ equals the multiplication operator $A(x)$ given by \eqref{Axgammadelta}.)

\begin{proof}[Derivation of Assertion~\ref{ass:unitary_HS_evolution}]
From \eqref{HdIdef} it is clear that $\Hd_I(x)$ is Hermitian (i.e., formally self-adjoint). From the expression \eqref{HdIxSigma} and the CCR/CAR, it is clear that
\be
[\Hd_I(x),\Hd_I(x')]=0
\ee
for $x\neq x'\in\Sigma$, and thus for any spacelike separated $x$ and $x'$. Thus, the Tomonaga--Schwinger equation defines a unitary operator $\tilde{U}_{\Sigma\to\Sigma'}:\tilde\Hilbert\to\tilde\Hilbert$, and the full time evolution 
\be
U_{\Sigma\to\Sigma'} = F_{\Sigma_0\to\Sigma'}\tilde{U}_{\Sigma\to\Sigma'}F_{\Sigma\to\Sigma_0}
\ee
is a unitary isomorphism $U_{\Sigma\to\Sigma'}:\Hilbert_\Sigma\to\Hilbert_{\Sigma'}$ because the $F$ are unitary isomorphisms.
\end{proof}

\label{proofs:TStophi}

\begin{proof}[Derivation of Assertion~\ref{ass:TStophi}]
Fix a configuration $q$ and consider two spacelike hypersurfaces $\Sigma,\Sigma'$ with $q\subset \Sigma\cap \Sigma'$. In order to derive \eqref{psiSigmaprimeSigma}, it suffices to consider the case that $\Sigma,\Sigma'$ are infinitesimally neighboring, as that will imply \eqref{psiSigmaprimeSigma} also for a finite difference between $\Sigma$ and $\Sigma'$. So the Tomonaga--Schwinger equation says
\be
i(\tilde{\psi}_{\Sigma'}-\tilde{\psi}_\Sigma) = \biggl( \int_\Sigma^{\Sigma'} d^4x\, \Hd_I(x) \biggr)\tilde\psi_\Sigma
\ee
and therefore
\be
i(F_{\Sigma'\to\Sigma}\psi_{\Sigma'}-\psi_\Sigma) = \biggl( \int_\Sigma^{\Sigma'} d^4x\, F_{\Sigma}\,\Hd_I(x)\, F_{\Sigma}^{-1} \biggr)\psi_\Sigma\,.
\ee
By \eqref{freepsiSigmaprimeSigma},
\be
F_{\Sigma'\to\Sigma}\psi_{\Sigma'}(q) = \psi_{\Sigma'}(q)\,,
\ee
so
\begin{align}
i\bigl(\psi_{\Sigma'}(q)-\psi_\Sigma(q) \bigr)
&= i\bigl(F_{\Sigma'\to\Sigma}\psi_{\Sigma'}(q)-\psi_\Sigma(q)\bigr)\\
&=\biggl( \int_\Sigma^{\Sigma'} d^4x\, F_{\Sigma}\,\Hd_I(x)\, F_{\Sigma}^{-1} \biggr)\psi_\Sigma(q)\\
&=d\ell\biggl(\int_\Sigma d^3x \, \tau(x) \, F_{\Sigma}\,\Hd_I(x)\, F_{\Sigma}^{-1} \biggr)\psi_\Sigma(q)\,.
\end{align}
It remains to check that the last expression vanishes. Note that those $\Hd_I(x)$ with $x\in q$ do not contribute because, for such $x$, $\tau(x)=0$. Now consider $\Hd_I(x)$ with $x\notin q$. To understand what happens with the continuous tensor product, it is easiest to think about a finite tensor product $\Hilbert=\Hilbert_1\otimes\Hilbert_2$; a configuration $q$ then may correspond to, say, having 0 particles at location 1 and some particles at location 2; to evaluate $\psi_\Sigma$ at a configuration $q$ then corresponds, according to \eqref{psiSigmaqdef}, to carrying out the partial inner product $\scp{\emptyset_1}{\Psi}$, where $|\emptyset_1\rangle\in\Hilbert_1$ and $\Psi\in\Hilbert$ plays the role of $\psi_\Sigma$. Now using \eqref{HdIxHilbertx}, $F_{\Sigma}\,\Hd_I(x)\, F_{\Sigma}^{-1}$ corresponds for $x\notin q$ to an operator of the form $T_1\otimes I_2$ with Hermitian $T_1$; using \eqref{HdIxHilbertx0}, we have that $T_1|\emptyset_1\rangle =0$. Now observe that 
\be
\scp{\emptyset_1}{T_1\otimes I_2|\Psi} = 0\,.
\ee
For the same reason,
\be
F_{\Sigma}\,\Hd_I(x)\, F_{\Sigma}^{-1} \psi_\Sigma(q)=0
\ee
for $x\notin q$.
\end{proof}

\subsection{Permutation Symmetry}
\label{sec:proof_permutation}

\begin{proof}[Derivation of Assertion~\ref{ass:permutation}]
Fix a configuration $(x^{4M},y^{4N})\in \sS_{xy}$, and let $\Sigma$ be a spacelike hypersurface such that all $x_1,\ldots,x_M,y_1,\ldots,y_N\in\Sigma$. To establish \eqref{permutation}, it suffices to show that $\psi_\Sigma$ has the usual permutation symmetry (fermionic in the $x$-particles and bosonic in the $y$-particles). Since the free time evolution $F_{\Sigma\to\Sigma'}$ preserves permutation symmetry, it suffices that $\tilde\psi_\Sigma$ has the usual symmetry. By Assertion~\ref{ass:Tomonaga-Schwinger}, and since the Tomonaga--Schwinger equation with interaction Hamiltonian \eqref{HdIdef2} preserves permutation symmetry, this is the case, provided the initial datum has the usual symmetry.
\end{proof}

\subsection{Operator-Valued Fields}
\label{proofs:Heisenberg_Picture}

We now turn to Assertion~\ref{ass:Heisenberg_Picture}; first, we need another auxiliary assertion.

\begin{ass}\label{ass:ax}
For the emission--absorption model and any spacelike hypersurface $\Sigma$ containing $x$,
\be\label{aaSigma}
a_{r}(x)= U_{\Sigma\to\Sigma_0} a_{\Sigma,r}(x) U_{\Sigma_0\to\Sigma}\,,
\ee
where $U_{\Sigma\to\Sigma'}$ denotes the full time evolution. Likewise for $a^\dagger_r(x),b_s(y)$, and $b^\dagger_s(y)$. As a consequence, the CAR/CCR also hold\footnote{That is, \eqref{CARCCRfirst}--\eqref{CARCCRlast} are still valid if we drop the index $\Sigma$ on $a$'s and $b$'s. The delta function, however, now understood as $U_{\Sigma\to\Sigma_0} \,\delta^3_\Sigma(x-x') \, U_{\Sigma_0\to \Sigma}$, keeps the index $\Sigma$, with $\Sigma$ an arbitrary spacelike hypersurface.} for the $a^{\#}_r(x)$ and $b^\circ_s(y)$.
\end{ass}

\begin{proof}[Derivation]
Equation~\eqref{aaSigma} is obviously true for $\Sigma=\Sigma_t$ with $t=x^0$. Now consider an arbitrary $\Sigma$ containing $x$, and consider a continuous family of spacelike hypersurfaces $(\Sigma^{s})_{s\in [0,1]}$ interpolating between $\Sigma^{0}=\Sigma_t$ and $\Sigma^{1}=\Sigma$ and satisfying $x\in\Sigma^{s}$. We will show that if \eqref{aaSigma} is true of $\Sigma^{s}$ then it is also true of $\Sigma^{s+ds}$; we know it is true of $\Sigma^{0}$ and will conclude that it is true of $\Sigma^{1}$. From the Tomonaga--Schwinger equation, we obtain that
\be
\tilde U_{\Sigma^{s}\to\Sigma^{s+ds}} = I -i\int_{\Sigma^{s}}^{\Sigma^{s+ds}} d^4y \, \Hd_I(y)= I-i\, ds \int_{\Sigma^s} d^3y \, \tau(y)\, \Hd_I(y)\,,
\ee
where $\tau(y)\, ds$ is the signed thickness of the layer between $\Sigma^s$ and $\Sigma^{s+ds}$. Thus, recalling \eqref{tildeUSigmaSigma'},
\be\label{USigmasSigmasds}
U_{\Sigma^{s}\to\Sigma^{s+ds}}  = F_{\Sigma^s\to\Sigma^{s+ds}}-i\, ds\, F_{\Sigma^s\to\Sigma^{s+ds}} \int_{\Sigma^s} d^3y \, \tau(y)\, F_{\Sigma_0\to\Sigma^s}\Hd_I(y)F_{\Sigma^s\to\Sigma_0}
\ee
and 
\begin{align}
&U_{\Sigma_0\to\Sigma^{s+ds}}a_r(x)U_{\Sigma^{s+ds}\to\Sigma_0} \nonumber\\
&\quad = U_{\Sigma^s\to\Sigma^{s+ds}} U_{\Sigma_0\to\Sigma^s} a_r(x) U_{\Sigma^s\to\Sigma_0} U_{\Sigma^{s+ds}\to\Sigma^s}\\
\intertext{[by the assumption that \eqref{aaSigma} is true of $\Sigma^s$]}
&\quad =  U_{\Sigma^s\to\Sigma^{s+ds}} a_{\Sigma^s,r}(x) U_{\Sigma^{s+ds}\to\Sigma^s}\\
\intertext{[by \eqref{aSigmaprimeSigmafree}]}
&\quad = U_{\Sigma^s\to\Sigma^{s+ds}} F_{\Sigma^{s+ds}\to\Sigma^s} a_{\Sigma^{s+ds},r}(x) F_{\Sigma^s\to\Sigma^{s+ds}} U_{\Sigma^{s+ds}\to\Sigma^s}\\
\intertext{[by \eqref{USigmasSigmasds}, to first order in $ds$]}
&\quad =a_{\Sigma^{s+ds},r}(x)+i\, ds\Biggl[a_{\Sigma^{s+ds},r}(x), \int_{\Sigma^s} d^3y \, \tau(y)\, F_{\Sigma_0\to\Sigma^{s+ds}}\Hd_I(y)F_{\Sigma^{s+ds}\to\Sigma_0}  \Biggr]\\
&\quad =a_{\Sigma^{s+ds},r}(x)+i\, ds\, F_{\Sigma^s\to\Sigma^{s+ds}}\Biggl[a_{\Sigma^{s},r}(x), \int_{\Sigma^s} d^3y \, \tau(y)\, F_{\Sigma_0\to\Sigma^{s}}\Hd_I(y)F_{\Sigma^{s}\to\Sigma_0}  \Biggr] F_{\Sigma^{s+ds}\to\Sigma^s}\,.\label{infinitesimal}
\end{align}
We show that the commutator vanishes. From \eqref{HdIxSigma} we have that, for $y\in\Sigma^s$,
\be
F_{\Sigma_0\to\Sigma^{s}}\Hd_I(y)F_{\Sigma^{s}\to\Sigma_0} = \sum_{r'=1}^4 a_{\Sigma^s,r'}^\dagger(y)a_{\Sigma^s,r'}(y) \otimes \sum_{s'=1}^4 \bigl(g_{s'}^* b_{\Sigma^s,s'}(y) + g_{s'} b^\dagger_{\Sigma^s,s'}(y) \bigr)\,. 
\ee
By the CAR \eqref{CARaadagger},
\begin{align}
&a_{\Sigma^s,r}(x) \, a^\dagger_{\Sigma^s,r'}(y) \, a_{\Sigma^s,r'}(y) - a^\dagger_{\Sigma^s,r'}(y)\, a_{\Sigma^s,r'}(y)\, a_{\Sigma^s,r}(x) \nonumber\\
&\quad=a_{\Sigma^s,r}(x) \, a^\dagger_{\Sigma^s,r'}(y) \, a_{\Sigma^s,r'}(y) + a^\dagger_{\Sigma^s,r'}(y) \,a_{\Sigma^s,r}(x) \,  a_{\Sigma^s,r'}(y)\nonumber\\
\label{applyCAR1}
&\quad -a^\dagger_{\Sigma^s,r'}(y) \,a_{\Sigma^s,r}(x) \,  a_{\Sigma^s,r'}(y)- a^\dagger_{\Sigma^s,r'}(y)\, a_{\Sigma^s,r'}(y)\, a_{\Sigma^s,r}(x) \\
&\quad=\bigl( \gamma^{\underline{0}} \gamma^0 \bigr)_{rr'} \,\delta^3_{\Sigma^s}(x-y)\,  a_{\Sigma^s,r'}(y)\,,
\label{applyCAR2}
\end{align}
so
\begin{align}
\Biggl[a_{\Sigma^s,r}(x), F_{\Sigma_0\to\Sigma^{s}}\Hd_I(y)F_{\Sigma^{s}\to\Sigma_0} \Biggr]
&= \sum_{r'}\bigl( \gamma^{\underline{0}} \gamma^0 \bigr)_{rr'} \,\delta^3_{\Sigma^s}(x-y)\, a_{\Sigma^s,r'}(y)\nonumber\\
&\quad \otimes \sum_{s'=1}^4 \bigl(g_{s'}^* b_{\Sigma^s,s'}(y) + g_{s'} b^\dagger_{\Sigma^s,s'}(y) \bigr) 
\end{align}
and thus, we find for the commutator in \eqref{infinitesimal},
\begin{align}
\Biggl[a_{\Sigma^s,r}(x), \int \Biggr]
&= \int_{\Sigma^s} d^3y \, \tau(y)\,\sum_{r'}\bigl( \gamma^{\underline{0}} \gamma^0 \bigr)_{rr'} \,\delta^3_{\Sigma^s}(x-y)\,  a_{\Sigma^s,r'}(y)\nonumber\\
&\quad \otimes \sum_{s'=1}^4 \bigl(g_{s'}^* b_{\Sigma^s,s'}(y) + g_{s'} b^\dagger_{\Sigma^s,s'}(y) \bigr) \\
&= \tau(x)\, \sum_{r'}\bigl( \gamma^{\underline{0}} \gamma^0 \bigr)_{rr'} \, a_{\Sigma^s,r'}(x)\nonumber\\
&\quad \otimes \sum_{s'=1}^4 \bigl(g_{s'}^* b_{\Sigma^s,s'}(x) + g_{s'} b^\dagger_{\Sigma^s,s'}(x) \bigr) \\
&=0
\end{align}
because $\tau(x)=0$. This completes the proof of \eqref{aaSigma}.

The only part of the reasoning that is different for $b_s(y)$ instead of $a_r(x)$ is the application in \eqref{applyCAR1} and \eqref{applyCAR2} of the CAR/CCR, which reads instead
\be\label{applyCCR}
\Bigl[ b_{\Sigma^s,s}(x), \bigl(g_{s'}^* b_{\Sigma^s,s'}(y) + g_{s'} b^\dagger_{\Sigma^s,s'}(y) \bigr)\Bigr]=
g_{s'}\, \bigl( \gamma^{\underline{0}} \gamma^0 \bigr)_{ss'} \, \delta^3_{\Sigma^s}(x-y)\,,
\ee
leading to
\be
\Biggl[b_{\Sigma^s,s}(x), F_{\Sigma_0\to\Sigma^{s}}\Hd_I(y)F_{\Sigma^{s}\to\Sigma_0} \Biggr]
=(\gamma^{\underline{0}} \gamma^0 g)_s\, \delta^3_{\Sigma^s}(x-y)\sum_{r'=1}^4 a_{\Sigma^s,r'}^\dagger(y)a_{\Sigma^s,r'}(y) 
\ee
and thus
\begin{align}
\Biggl[b_{\Sigma^s,s}(x), \int \Biggr]
&= \int_{\Sigma^s} d^3y \, \tau(y)\,
(\gamma^{\underline{0}} \gamma^0 g)_s \, \delta^3_{\Sigma^s}(x-y)\sum_{r'=1}^4 a_{\Sigma^s,r'}^\dagger(y)a_{\Sigma^s,r'}(y) \\
&= \tau(x)\, (\gamma^{\underline{0}} \gamma^0 g)_{s} \sum_{r'=1}^4 a_{\Sigma^s,r'}^\dagger(x)a_{\Sigma^s,r'}(x) \\
&=0
\end{align}
because $\tau(x)=0$.
\end{proof}

\begin{proof}[Derivation of Assertion~\ref{ass:Heisenberg_Picture}]
Choose a spacelike hypersurface $\Sigma$ such that $x_1,\ldots,x_M,y_1,\ldots, y_N\in\Sigma$, and let $U=U_{\Sigma_0\to\Sigma}$. Then $U|\emptyset\rangle=|\emptyset_\Sigma\rangle$ (because, for initial condition $\psi_0=|\emptyset\rangle$, $\phi(x^{4M},y^{4N})=0$ whenever $M>0$ or $N>0$); $U\psi_0=\psi_\Sigma$; by Assertion~\ref{ass:ax}, $U\, a_r(x) \, U^{-1}=a_{\Sigma,r}(x)$. Thus,
\begin{align}
&\scp{\emptyset}{a_{r_1}(x_1)\cdots a_{r_M}(x_M)b_{s_1}(y_1)\cdots b_{s_N}(y_N)|\psi_0}\nonumber\\
&\quad = \scp{U\emptyset}{Ua_{r_1}(x_1)U^{-1}\cdots Ua_{r_M}(x_M)U^{-1}Ub_{s_1}(y_1)U^{-1}\cdots U b_{s_N}(y_N)U^{-1}|U\psi_0}\\
&\quad = \scp{\emptyset_\Sigma}{a_{\Sigma,r_1}(x_1)\cdots a_{\Sigma,r_M}(x_M)b_{\Sigma,s_1}(y_1)\cdots  b_{\Sigma,s_N}(y_N)|\psi_\Sigma}_\Sigma\\
&\quad =\sqrt{M!N!}\, (-1)^{M(M-1)/2}\,\psi_{\Sigma,r_1\ldots r_M,s_1\ldots s_N}(x^{4M},y^{4N})\,.
\end{align}
This proves \eqref{phiabPsi}, also if some of the points coincide. 

To prove \eqref{phiFPsi} for pairwise distinct $x_j,y_k$, we use \eqref{Fabdef}, the CAR/CCR, and the facts that, for any $x,y\in\Sigma$,
\be\label{emptycreation0}
\langle\emptyset_\Sigma|a^\dagger_{\Sigma,r}(x) =0\,,\quad
\langle\emptyset_\Sigma|b^\dagger_{\Sigma,s}(y) =0\,.
\ee
Indeed, expanding the $\Phi_{x,r}(x_j)$ and $\Phi_{y,s}(y_k)$ in \eqref{phiFPsi}, we obtain $2^{M+N}$ terms, each of which is a product of $M$ factors $a^{\#}_{\Sigma,r_j}(x_j)$ and then $N$ factors $b^{\circ}_{\Sigma,s_k}(y_k)$. By the CAR, and since the $x_j$ are pairwise distinct, $a_{\Sigma,r}(x)a^\dagger_{\Sigma,r'}(x')=-a^\dagger_{\Sigma,r'}(x')a_{\Sigma,r}(x)$; that is, the creation operators $a^\dagger$ can be moved to the left at the expense of a minus sign. Likewise, every $b^\dagger$ can be moved to the left of the $b$ operators (without changing the sign), and since all $b^\circ$ commute with all $a^{\#}$, the $b^\dagger$ can in fact be moved to the very left of the product of $M+N$ factors. If there is any creation operator at all on the very left, then it will send $\langle\emptyset_\Sigma|$ to 0 according to \eqref{emptycreation0}. Thus, only one term out of the $2^{M+N}$ is nonzero: that which contains no creation operator $a^\dagger$ or $b^\dagger$. That term is exactly the expression in \eqref{phiabPsi}, and thus equal (up to the combinatorial factor and the sign) to $\phi$.
\end{proof}

\bigskip

\noindent{\it Acknowledgments.} We thank Detlef D\"urr, Felix Finster, Sheldon Goldstein, Michael Kiessling, Matthias Lienert, and Hrvoje Nikoli\'c for helpful discussions. S.P.\ acknowledges support from Cusanuswerk, from the German--American Fulbright Commission, and from the European Cooperation in Science and Technology (COST action MP1006). R.T.\ acknowledges support from the John Templeton Foundation (grant no.\ 37433) and from the Trustees Research Fellowship Program at Rutgers. 


\begin{thebibliography}{19}

\bibitem{BB}
I.~Bialynicki-Birula:
\newblock Photon Wave Function.
\newblock In E.~Wolf (ed.) \textit{Progress in Optics, Vol.\ XXXVI}.
\newblock Amsterdam: Elsevier (1996)

\bibitem{bloch:1934}
F.~Bloch:
\newblock {Die physikalische Bedeutung mehrerer Zeiten in der
  Quantenelektrodynamik}.
\newblock {\em Physikalische Zeitschrift der Sowjetunion}, 5:301--305 (1934)

\bibitem{dimock:1982} 
J.~Dimock:
\newblock Dirac Quantum Fields on a Manifold.
\newblock \textit{Transactions AMS}, 269:133--147 (1982)

\bibitem{dirac:1932}
P.~A.~M. Dirac:
\newblock {Relativistic Quantum Mechanics}.
\newblock {\em Proceedings of the Royal Society London A}, 136:453--464 (1932)

\bibitem{dfp:1932}
P.~A.~M.~Dirac, V.~A.~Fock, and B.~Podolsky:
\newblock {On Quantum Electrodynamics}.
\newblock {\em Physikalische Zeitschrift der Sowjetunion}, 2(6):468--479 (1932).
\newblock Reprinted in J. Schwinger: {\em Selected Papers on Quantum
  Electrodynamics}, New York: Dover (1958)

\bibitem{DV82b} 
Ph.~Droz-Vincent:  
\newblock Second quantization of directly interacting particles. 
\newblock Pages 81--101 in J.~Llosa (ed.): \textit{Relativistic Action at a Distance: Classical and Quantum Aspects}, Berlin: Springer-Verlag (1982)

\bibitem{DV85} 
Ph.~Droz-Vincent: 
\newblock Relativistic quantum mechanics with non conserved number of particles. 
\newblock \textit{Journal of Geometry and Physics}, 2(1):101--119 (1985)

\bibitem{Fey65} R.~P.~Feynman: Nobel lecture (1965) 
\newblock \url{http://www.nobelprize.org/nobel_prizes/physics/laureates/1965/feynman-lecture.html}

\bibitem{Fin13}
F.~Finster:
Perturbative Quantum Field Theory in the Framework of the Fermionic Projector.
Preprint (2013)
\url{http://arxiv.org/abs/1310.4121}

\bibitem{GTTZ:2013}
S.~Goldstein, J.~Taylor, R.~Tumulka, and N.~Zangh\`\i:
\newblock The Fermionic Line Bundle.
\newblock Preprint (2014)

\bibitem{Gue52}
M.~G\"unther:
\newblock The Relativistic Configuration Space Formulation of the Multi-Electron Problem.
\newblock \textit{Physical Review}, 88(6): 1411--1421 (1952)

\bibitem{KTT:1947a}
Z.~Koba, T.~Tati, and S.~Tomonaga:
\newblock On a Relativistically Invariant Formulation of the Quantum Theory of
	Wave Fields. II. Case of Interacting Electromagnetic and Electron Fields.
\newblock {\em Progress of Theoretical Physics}, 2(3):101--116 (1947)

\bibitem{KTT:1947b}
Z.~Koba, T.~Tati, and S.~Tomonaga:
\newblock On a Relativistically Invariant Formulation of the Quantum Theory of
	Wave Fields. III. Case of Interacting Electromagnetic and Electron Fields.
\newblock {\em Progress of Theoretical Physics}, 2(4):198--208 (1947)

\bibitem{LP:1930}
L.~Landau and R.~Peierls:
\newblock Quantenelektrodynamik im Konfigurationsraum.
\newblock {\em Zeitschrift f\"ur Physik}, 62(3-4):188--200 (1930)

\bibitem{Marx:1972}
E.~Marx:
\newblock Generalized Relativistic Fock Space.
\newblock \textit{International Journal of Theoretical Physics}, 6(5):359--363 (1972)

\bibitem{Nel64} E.~Nelson: 
\newblock Interaction of Nonrelativistic Particles with a Quantized Scalar Field.
\newblock \textit{Journal of Mathematical Physics}, 5:1190--1197 (1964)

\bibitem{Nik10} H.~Nikoli\'c:
\newblock QFT as pilot-wave theory of particle creation and destruction.
\newblock \textit{International Journal of Modern Physics A}, 25:1477--1505 (2010)
\newblock \url{http://arxiv.org/abs/0904.2287}

\bibitem{PR84}
R.~Penrose and W.~Rindler:
\textit{Spinors and space-time. Vol.~1: Two-spinor calculus and relativistic fields.}
Cambridge University Press (1984)

\bibitem{pt:2013a}
S.~Petrat and R.~Tumulka:
\newblock Multi-Time Schr\"odinger Equations Cannot Contain Interaction Potentials.
\newblock Preprint (2013)
\newblock \url{http://arxiv.org/abs/1308.1065}

\bibitem{pt:2013e}
S.~Petrat and R.~Tumulka:
\newblock Multi-Time Equations, Classical and Quantum.
\newblock To appear in \textit{Proceedings of the Royal Society A} (2014)
\newblock \url{http://arxiv.org/abs/1309.1103}

\bibitem{pt:2013d}
S.~Petrat and R.~Tumulka:
\newblock Multi-Time Formulation of Pair Creation.
\newblock Preprint (2014)
\newblock \url{http://arxiv.org/abs/1401.6093}

\bibitem{pt:2013b}
S.~Petrat and R.~Tumulka:
\newblock Consistency of Evolution Equations for Multi-Time Wave Functions.
\newblock In preparation (2014)

\bibitem{RS2} 
M.~Reed and B.~Simon: 
\newblock \textit{Methods of Modern Mathematical Physics. Vol. 2:
	Fourier Analysis, Self-Adjointness.}
\newblock San Diego: Academic Press (1975)

\bibitem{schweber:1961}
S.~Schweber:
\newblock {\em {An Introduction To Relativistic Quantum Field Theory}}.
\newblock Row, Peterson and Company (1961)

\bibitem{schwinger:1948}
J.~Schwinger:
\newblock Quantum Electrodynamics. I. A Covariant Formulation.
\newblock {\em Physical Review} 74(10):1439--1461 (1948)

\bibitem{SZ}
M.~O.~Scully and M.~S.~Zubairy:
\newblock {\em Quantum Optics}.
\newblock Cambridge University Press (1997)

\bibitem{thaller:1992} 
B.~Thaller: 
\newblock \textit{The Dirac Equation}. 
\newblock Berlin: Springer-Verlag (1992)

\bibitem{tomonaga:1946}
S.~Tomonaga:
\newblock {On a Relativistically Invariant Formulation of the Quantum Theory of
  Wave Fields}.
\newblock {\em Progress of Theoretical Physics}, 1(2):27--42 (1946)

\bibitem{WG64} A.~S.~Wightman and L.~G\aa rding:
\newblock Fields as Operator-Valued Distributions.
\newblock \textit{Arkiv f\"or Fysik}, 28: 129--189 (1964)

\end{thebibliography}
\end{document}